\documentclass[
reprint,
superscriptaddress,
nobibnotes,
nofootinbib,
amsmath,amssymb,
aps, pra
]{revtex4-2}

\usepackage{graphicx}
\usepackage[dvipsnames]{xcolor} 
\usepackage{dcolumn}
\usepackage{braket}%
\usepackage{bm}
\usepackage[%
  pdftex,
  breaklinks=true,
  colorlinks=true,
  linkcolor=blue,
  urlcolor=blue,
  citecolor=blue
]{hyperref} 

\newcommand{\ad}{\hat{a}^{\text{\textdagger}}}

\newcommand{\la}{\left\langle}
\newcommand{\ra}{\right\rangle}

\newcommand{\rla}{\right\rangle \left\langle}

\newcommand{\gmin} {g_{\text{min}}}

\begin{document}

\title{Observation of Genuine Tripartite Non-Gaussian Entanglement from a Superconducting Three-Photon Spontaneous Parametric Down-Conversion Source}

\author{Benjamin~Jarvis-Frain\textsuperscript{*\textdagger}}
\affiliation{Institute for Quantum Computing and Department of Electrical and Computer Engineering, University of Waterloo,
Waterloo, Ontario N2L 3G1, Canada}
\author{Andy~Schang\textsuperscript{*}}
\affiliation{Institute for Quantum Computing and Department of Electrical and Computer Engineering, University of Waterloo,
Waterloo, Ontario N2L 3G1, Canada}
\author{Fernando Quijandr\'{\i}a} 
\affiliation{RIKEN Center for Quantum Computing, RIKEN, Wakoshi, Saitama 351-0198, Japan}
\author{Ibrahim~Nsanzineza}
\affiliation{Institute for Quantum Computing and Department of Electrical and Computer Engineering, University of Waterloo,
Waterloo, Ontario N2L 3G1, Canada}
\author{Dmytro~Dubyna}
\affiliation{Institute for Quantum Computing and Department of Electrical and Computer Engineering, University of Waterloo,
Waterloo, Ontario N2L 3G1, Canada}
\author{C. W.~Sandbo~Chang}
\affiliation{Institute for Quantum Computing and Department of Electrical and Computer Engineering, University of Waterloo,
Waterloo, Ontario N2L 3G1, Canada}
\affiliation{RIKEN Center for Quantum Computing, RIKEN, Wakoshi, Saitama 351-0198, Japan}
\author{Franco~Nori} 
\affiliation{RIKEN Center for Quantum Computing, RIKEN, Wakoshi, Saitama 351-0198, Japan}
\affiliation{Physics Department, University of Michigan, Ann Arbor, MI 48109-1040, USA}
\author{C.M.~Wilson\textsuperscript{\textdaggerdbl}}
\affiliation{Institute for Quantum Computing and Department of Electrical and Computer Engineering, University of Waterloo,
Waterloo, Ontario N2L 3G1, Canada}
\footnotetext[1]{These authors contributed equally to this work.}
\footnotetext[2]{Present address: Rigetti Computing, Berkeley, CA 94710, USA}
\footnotetext[3]{chris.wilson@uwaterloo.ca}

\date{\today}

\begin{abstract}
The generation of entangled photons through Spontaneous Parametric Down-Conversion (SPDC) is a critical resource for many key experiments and technologies in the domain of quantum optics. Historically, SPDC was limited to the generation of photon pairs. However, the use of the strong nonlinearities in circuit quantum electrodynamics has recently enabled the observation of Three-Photon SPDC (3P-SPDC). Despite great interest in the entanglement structure of the resultant states, entanglement between photon triplets produced by a 3P-SPDC source has still has not been confirmed. Here, we report on the observation of genuine tripartite non-Gaussian entanglement in the steady-state output field of a 3P-SPDC source consisting of a superconducting parametric cavity coupled to a transmission line. We study this non-Gaussian tripartite entanglement using an entanglement witness built from three-mode correlation functions, observing a maximum violation of the bound by 15 standard deviations of the statistical noise. Furthermore, we find strong agreement between the observed 
and the analytically predicted scaling of the entanglement witness. 
We then explore the impact of the temporal function used to define the photon mode on the observed value of the entanglement witness.

\end{abstract}

\maketitle

The ability to generate entangled pairs of photons via spontaneous parametric down-conversion (SPDC)~\cite{SPDC_FirstDemonstration,Scully_Zubairy_1997} has enabled many achievements in fundamental quantum information science~\cite{SPDC_Bell_ParAmp, SPDC_Bell_10km, EarlyHOMExperiment} and its applications~\cite{AttosecondHOM, SPDC_QKD, ShorWithSPDC}. The natural extension of this phenomenon to  generating photon triplets from Three-Photon SPDC (3P-SPDC) was studied theoretically starting over 30 years ago~\cite{GeneralizedSqueezing, banaszek&knight}. However, 3P-SPDC was only recently demonstrated using superconducting circuits operating at microwave frequencies~\cite{scw_chang2020, cubicphase}. This development has spurred renewed interest in the phenomenon and its potential applications~\cite{Fisher_3Photon,3rdOrder_ThermalPumping_Degenerate,NonG_Ent_Bencheikh1,NonG_Ent_Bencheikh2,NonG_Ent_Bencheikh3,From3PSPDC_toCats,MetrologyOfNonG_Ent,NonG_Steering,PumpOn_Ent,1j44-q3bw,Banic_2022}.
These theoretical studies predict a variety of useful nonclassical states, depending on the degeneracy of the modes involved. In the case of degenerate 3P-SPDC into a single mode, the resulting trisqueezed state was predicted to exhibit Wigner negativity~\cite{banaszek&knight} and can be used to generate approximate cubic phase states, which are themselves non-Gaussian resource states for universal continuous-variable quantum computation~\cite{Gaussian_CPG, cubicphase}. Alternatively, in the case of nondegenerate down-conversion into three different modes, the photons have been theoretically shown to exhibit genuine tripartite non-Gaussian entanglement~\cite{Our_condition, NonG_Ent_Bencheikh3} and can, for example, be used as a source of heralded two-photon entangled states~\cite{3(!S)PDC_HeraldPairs}. These results on 3P-SPDC add to a growing body of work on nonclassical effects in few-photon nonlinear optics~\cite{PhysRevLett.117.043601,PhysRevA.95.063849,Stassi_2017,Stassi_2016,Kockum2017} including recent experimental demonstrations~\cite{nature_1_to_2photon_ent,Tomonaga2025, Roch_Downconversion}.

In this letter, we study the tripartite entanglement of nondegenerate 3P-SPDC into three different modes. In Ref.~\cite{3P_NotEntangled} it was demonstrated that it is impossible to detect tripartite entanglement from 3P-SPDC using only moments of the field quadratures, $\hat{X}$ and $\hat{P}$,  up to second order. That is, entanglement in these states cannot be detected using standard measures developed for Gaussian states. Subsequently, it was shown that the entanglement of these states could be detected with higher moments of the quadratures.
Consequently, the witness we use to observe this tripartite entanglement is built from moments up to fourth order~\cite{Our_condition}.

Since our entanglement witness depends on the third moments of the quadratures, it certifies that the states found to be entangled are also non-Gaussian (given that the states are mean-zero)~\cite{supplementary, Agarwal_2012,CVQC}. Non-Gaussian entanglement has recently been of great interest in the field of quantum information due to its advantages over Gaussian entanglement in tasks like sensing and metrology~\cite{QI_with_photon_sub,QI_with_cond_meas,3P_Illumination,Fisher_3Photon}. 
Furthermore, non-Gaussian entanglement has been shown to be necessary for quantum advantage in many bosonic 
quantum computational schemes~\cite{BosonicQResources}. For instance, in the continuous-variable cluster state scheme, particular non-Gaussian states are a sufficient resource along with a complete set of Gaussian operations for universal quantum computation~\cite{cluster_states,CPS_to_CPG}.

\begin{figure}
    \centering
    \includegraphics[scale = 0.66]{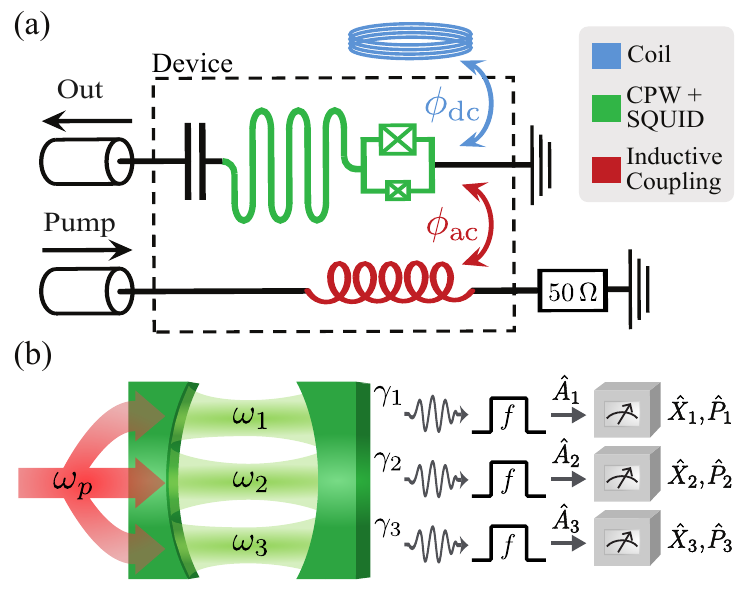}
    \caption{Three-photon spontaneous parametric down-conversion (3P-SPDC) in a superconducting cavity. (a) Circuit model. The short circuit at one end of a meandered $\lambda/4$  CPW resonator (green) is replaced by an asymmetric SQUID that creates a tunable boundary condition for the resonator modes. The SQUID is flux-coupled to a static DC coil (blue) and an AC pump line (red). The other end of the cavity is capacitively coupled to an output transmission line. (b) Cartoon of 3P-SPDC in the cavity with heterodyne measurement of the three propagating output modes defined by a temporal mode function. High-frequency pump photons (red) are split into entangled photon triplets inside the cavity at distinct frequencies (green) which couple to the continuum of the output line with coupling strengths ${\gamma_i}$. We convert the propagating modes into discrete modes, ${\hat{A}_i}$, with a temporal mode function $f(t)$,  lastly measuring their voltage quadratures ${\hat{X}_i}$ and ${\hat{P}_i}$.}
    \label{fig:3SPDC_Model_Figure}
\end{figure}

We implement nondegenerate 3P-SPDC inside a superconducting microwave cavity consisting of a coplanar waveguide (CPW) terminated by an asymmetric SQUID at one end of the resonator and capacitively coupled to a transmission line at the other, as illustrated in Fig.~\ref{fig:3SPDC_Model_Figure} (a)~\cite{DCE,Pumpistor,multimodeJPA}. The CPW constituting the resonator is approximately 48~mm long, supporting several modes over our 4-12~GHz measurement bandwidth. We vary the impedance of the line along its length such that the mode frequencies are not equally spaced~\cite{Sandbo_Gauss_3PEnt,CreutzLadder,creutz_theory,scw_chang2020}. The asymmetric SQUID is inductively coupled to an on-chip microwave pump line, which allows for a time-varying modulation of the flux $\phi_\text{ac}(t)$, as well as to an off-chip coil, which provides a static flux bias $\phi_\text{dc}$. 

The SQUID allows for the flux potential of the cavity to be parametrically modulated in time using the pump line. Furthermore, it has been shown that the asymmetry in the SQUID junctions, together with the correct frequency modulation, leads to the activation of odd-order terms in the Hamiltonian of the system~\cite{scw_chang2020, SvenssonPeriodMultiplication, SvenssonPeriodTripling}. The flux across the SQUID at the cavity termination is related to the bosonic operators of the cavity modes by  $\hat{\Phi}=\sum_i\Phi_i(\hat{a}_i+\ad_i)$, where $\Phi_i$ is the scale of zero-point fluctuations of the flux of the $i_\text{th}$ mode across the SQUID. The junction potential can be written as $\hat{H}_{\rm even}(\hat{\phi}, t) + \hat{H}_{\rm odd}(\hat{\phi}, t)$, where $\phi=2 \pi \Phi/\Phi_0$ and $\Phi_0$ is the flux quantum. Here, $\hat{H}_{\rm even}$ is the potential of a symmetric SQUID $\hat{H}_{\rm even}(\hat{\phi}, t) = -E_+ \cos [\phi_\text{ext}(t)] \cos \hat{\phi}$, with $E_+$ the sum of the Josephson energies of both junctions and $\phi_\text{ext}(t) = \phi_{\rm dc} + \phi_{\rm ac}(t)$ is the total external flux threading the SQUID. The contribution due to the junction asymmetry is $\hat{H}_{\rm odd} = -E_-\sin [\phi_\text{ext}(t)]\sin \hat{\phi}$, where $E_-$ is the difference between the junctions' Josephson energies. The lowest nonlinear contribution to the odd terms is $\hat{H}_{\rm odd} \propto \phi_{\rm ac}(t) E_- \sin \hat{\phi}$. Restricting to just three resonant modes of the cavity,
the Hamiltonian term accounting for the three-photon interaction is
\begin{align}
 \hat{H}_{\rm odd}=&\hbar g_0\left(\alpha e^{i\omega_p t}+\alpha^*e^{-i\omega_p t}\right)\nonumber\\&\left(\hat{a}
 _1+\ad_1\right)\left(\hat{a}_2+\ad_2\right)\left(\hat{a}_3+\ad_3\right),
\end{align}
where $g_0$ is a coupling constant dictated by the SQUID asymmetry as well as $\phi_{\rm dc}$, and $\alpha$ is the complex amplitude of the pump field. Finally, as is typical in circuit QED setups, there are residual Kerr interactions between the modes due to the time-independent part of $\hat{H}_{\rm even}$. 

By pumping at the sum frequency of the three cavity modes, that is  $\omega_p=\omega_1+\omega_2+\omega_3$, 
we obtain the following total Hamiltonian in an interaction frame after using a rotating-wave approximation,
\begin{align}\label{system_hamiltonian}
 \hat{H} &=\hbar g\left(\hat{a}_1\hat{a}_2\hat{a}_3+\ad_1\ad_2\ad_3\right) \nonumber\\ 
 &- \sum_{n, m}\chi_{nm}\ad_n\ad_m\hat{a}_n\hat{a}_m ,
\end{align}
where we have included the Kerr terms with strengths $\chi_{nm}$. In addition, $g = g_0\alpha$ is the effective three-photon interaction strength in the parametric approximation with our attention restricted to real values of $\alpha$.

A witness that could observe the tripartite entanglement of states generated by 3P-SPDC was proposed in Ref.~\cite{Our_condition}. 
This witness follows from Cauchy-Schwarz derived inequalities introduced in Refs.~\cite{Hillery2006,Mult_NonG_Fundamental}.
A similar witness was derived earlier in ~\cite{Miranowicz2009} by generalizing the method of matrices of moments of field operators 
introduced in~\cite{Shchukin2006}, and further extended in~\cite{Miranowicz2010}.  
Subsequently, other witnesses have been proposed, all of which use moments of the field operators of third order or higher~\cite{Fisher_3Photon,NonG_Steering,NonG_Ent_Bencheikh1}. The witnesses described in~\cite{Fisher_3Photon} and~\cite{NonG_Steering} detect entanglement, respectively, in terms of the quantum Fisher information and the capacity for quantum steering, both of which are appealing in that they measure entanglement through quantities directly connected to practical utility. However, we operate in the few-photon regime, where the thermal background is small but not negligible. 
For these reasons, we restrict to the original witness for genuine tripartite entanglement ~\cite{Our_condition}
\begin{equation}\label{W}
    W=|\langle \hat{a}_1\hat{a}_2\hat{a}_3\rangle|-\sum_{\substack{i,j,k=1,2,3 \\ i\neq j\neq k\neq i}}
    \sqrt{\la\hat{n}_i\rla \hat{n}_j\hat{n}_k\ra}, 
\end{equation}where $\hat{n}_i = \hat{a}_i^{\dagger}\hat{a}_i$ is the number operator on the $i_\text{th}$ mode. We note that this witness certifies both the weaker \textit{full inseparability} and stronger \textit{genuine multipartite entanglement}. For a detailed discussion of the distinction, see, e.g., Ref.~\cite{Our_condition}.

\begin{figure}[t]
\includegraphics[scale=0.35]{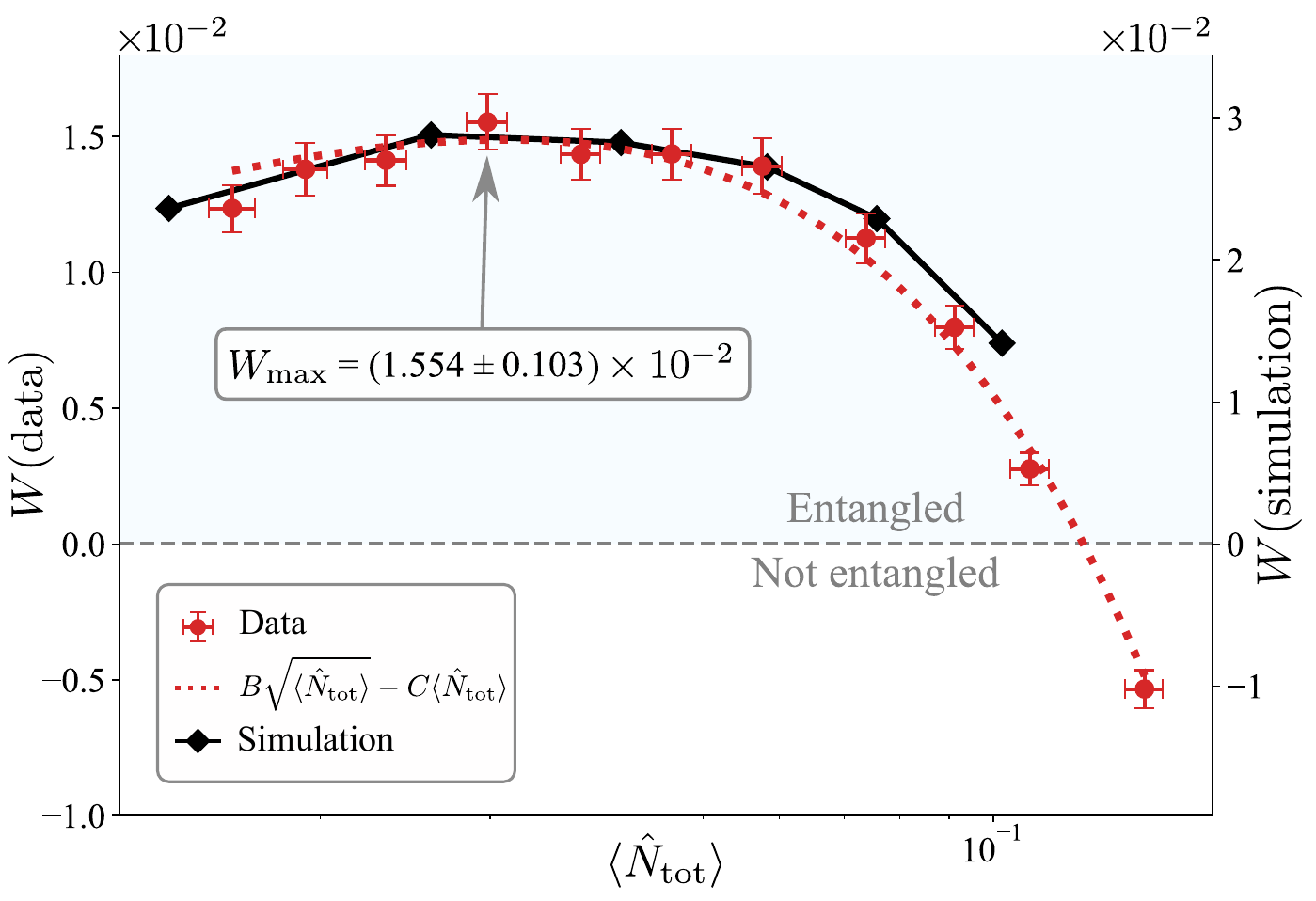}
\caption{\label{fig:epsart} Measured entanglement witness, $W$, for three-mode states generated from 3P-SPDC as a function of their measured average photon number, $\langle \hat{N}_{\textrm{tot}} \rangle$, which is varied by changing the pump amplitude. The error bars show one standard deviation of statistical error, $\sigma$, above and below the mean. We observe $W>0$, indicating genuine tripartite non-Gaussian entanglement, over nearly an order of magnitude of $\langle \hat{N}_{\textrm{tot}} \rangle$ before entanglement is no longer detected. At the maximum value of $W$, we violate the bound by more than $15\sigma$. We compare the measured witness values against stochastic trajectories simulations and an analytically predicted scaling law, Eq.~\ref{witness_form}. Simulation results are plotted on the right axis, rescaled by a fitted factor of 1.9.}
\end{figure}

Experimentally, the entangled photon triplets generated from 3P-SPDC inside the cavity leak into the capacitively coupled transmission line, where we measure them using a linear amplifier as illustrated in Fig. \ref{fig:3SPDC_Model_Figure}(b). Following digitization, we compute moments of the quadrature voltages of the propagating states~\cite{LinearDetection1}. In this work, we focus on measuring the entanglement directly in the propagating states, as opposed to reconstructing the entanglement in the cavity. In addition to being more experimentally accessible, tripartite entanglement in propagating states may also be a more practical form of nonclassical resource for applications like quantum-enhanced metrology and remote sensing~\cite{Fisher_3Photon, MetrologyOfNonG_Ent}.

In general, the propagating states in the transmission line are described by continuum operators, e.g., in the time domain as $\hat{a}_{\text{out},i}(t)$, where $i$ denotes the corresponding cavity mode. To analyze the correlations in the system, we reduce the continuum operators to single-mode operators, by defining appropriate mode functions~\cite{Loudon1990,Raymer_2020}. In particular, we use a time domain description, as spectral analysis of higher-order quantum correlation functions (beyond the spectral density) is not well developed. 
The discrete bosonic operator annihilating a photon occupying the mode defined by the function $f(t)$ is $\hat{A}_i  =\int \text{d}t\, f(t)\hat{a}_{\text{out},i}(t)$.
The normalization condition $\int_{0}^{\infty} {\rm d}t \, \vert f(t) \vert^2 = 1$ guarantees that the discretized modes obey the bosonic commutation relation $[\hat{A}_i,\hat{A}_j^\dagger]= \delta_{ij}$.
Here we assume that the three modes are being filtered with the same function $f(t)$. Although our RF digitizers have built-in antialiasing filters, we can control the mode function used in the experiments by oversampling the data and then digitally filtering with the chosen mode function.
In this work we investigate the steady-state emission from the continuously driven cavity by only filtering the output emission once the cavity field has reached a stationary regime.  

In our measurements, the propagating states are first amplified by a Crescendo traveling-wave parametric amplifier (TWPA) from Quantware, followed by a sequence of HEMT amplifiers. They are then measured using heterodyne detection with three room-temperature RF digitizers. The image band of each digitizer is rejected using an external superheterodyne mixing stage on each channel. The moments of the voltage quadratures of the states as they leave the cavity are reconstructed by calibrating the gain and noise of the amplifier chain using a shot noise tunnel junction provided by NIST Boulder. We measure the moments of the amplifier noise by taking interleaved measurements of the output signals with the down-conversion process on and off. Finally, we calibrate the measurement and parametric pump frequencies to account for the external flux dependence and Kerr terms in the SQUID potential that shift the cavity mode frequencies with pump power~\cite{supplementary}. 

\begin{figure}
    \centering
    \includegraphics[width=1\linewidth]{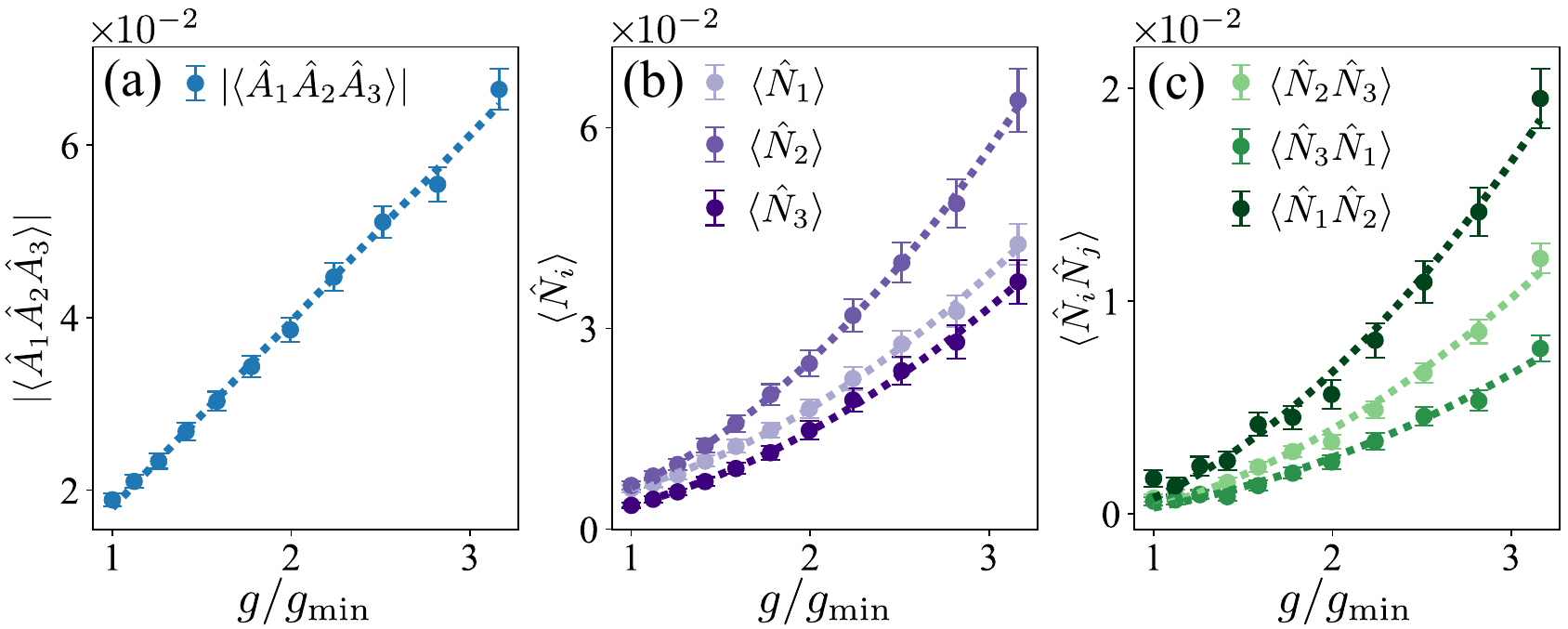}
    \caption{Measured correlators as a function of the normalized three-photon drive strength, $g/\gmin$. We estimate $\gmin = 0.057$~MHz~\cite{supplementary}. Dotted lines are the linear (a) and quadratic (b),(c) scalings of the correlators predicted by perturbation theory in equations (\ref{pt_1}-\ref{pt_3}) with fitted multiplicative factors.}
    \label{fig:Correlators}
\end{figure}

In Fig.~\ref{fig:epsart}, we plot the measured value of $W$ as a function of the total average number of photons, $\langle \hat{N}_{\rm tot} \rangle$, which is controlled by the pump strength. The temporal mode is fixed by a 1~MHz antialiasing filter built into the digitizers' firmware which corresponds to a windowed sinc function in the time domain~\cite{supplementary}. Entanglement is observed over nearly an order of magnitude of $\langle \hat{N}_{\rm tot} \rangle$, with an observed maximum value of $W=\left(1.554\pm0.103\right)\times 10^{-2}$, violating the entanglement bound by 15 standard deviations of the statistical noise.

To understand the behavior of $W$, we show the measured filtered output field correlators which contribute to the entanglement witness as a function of the normalized three-photon drive strength $g/g_{\text{min}}$ in Fig.~\ref{fig:Correlators}. The measured $\langle \hat{N}_{\rm tot} \rangle \lesssim 0.2$ 
place us in a weak driving regime which can be studied analytically using perturbation theory. As detailed in~\cite{supplementary}, we treat the three-photon Hamiltonian as a small perturbation to the undriven dissipative dynamics which cools the three mode cavity system to a thermal state close to the photon vacuum $\hat{\rho}_{\rm ss}^{(0)} \approx \ket{000}\bra{000}$. 
Our perturbative parameter is $\lambda = g /\gamma_T$, where $\gamma_T$ is the sum of the total decay rates of the three cavity modes, $\gamma_T= \sum_i \gamma_i$. We neglect Kerr nonlinearities in our analysis as they have minimal effect in the few photon regime. 
Up to second-order in $\lambda$, the cavity steady-state is expressed as $\hat{\rho}_{\rm ss} \approx \hat{\rho}_{\rm ss}^{(0)} + \lambda \hat{\rho}_{\rm ss}^{(1)} + \lambda^2 \hat{\rho}_{\rm ss}^{(2)}$. The first-order correction to the steady-state $\hat{\rho}^{(1)}_{\rm ss} \propto \lambda \ket{111}\bra{000} + {\rm h.c.}$ is solely responsible for the three-photon correlations $\langle \hat{a}_1 \hat{a}_2 \hat{a}_3 \rangle$, while terms contributing to the modes' populations $\langle \hat{n}_i \rangle$ and number-number correlations $\langle \hat{n}_i \hat{n}_j \rangle$ only appear in the second-order correction to the steady-state. The explicit relations for the lowest-order contributions to the steady-state cavity field correlations are
\begin{align}
    \vert \langle \hat{a}_1 \hat{a}_2 \hat{a}_3 \rangle \vert &= \frac{2g}{\gamma_T}\label{pt_1}\\
    \langle \hat{n}_i \rangle &= \left( \frac{2g}{\gamma_T} \right)^2 \left( \frac{\gamma_T}{\gamma_i} \right)\label{pt_2} \\
    \langle \hat{n}_i \hat{n}_j \rangle &= \left( \frac{2g}{\gamma_T} \right)^2 \left( \frac{\gamma_T}{\gamma_i + \gamma_j} \right) \label{pt_3}.
\end{align}
Starting from our steady-state approximation it is possible to calculate different order multitime correlations and, from these, the output field correlations. As shown in~\cite{supplementary}, the above 
leading order corrections to the different correlators map to the output field, 
with the explicit relations depending on the particular temporal mode profile $f(t)$.
This agrees with the scaling observed in Fig.~\ref{fig:Correlators}, that is
$\vert \langle \hat{A}_1\hat{A}_2\hat{A}_3 \rangle \vert\propto g$, while $\langle \hat{N}_i \rangle ,\, \langle \hat{N}_i \hat{N}_j \rangle \propto g^2$. Owing to these scaling laws and the definition of the entanglement witness~\eqref{W}, in the weak driving regime we have that $W \propto g$. Therefore, independent of the total decay rate, an arbitrarily weak but nonzero value of $g$ suffices to generate genuine tripartite entanglement in our system, which agrees with what we observe in Fig.~\ref{fig:epsart}. 
From the scalings of these correlators, we can predict a scaling law for $W$:
\begin{align}\label{witness_form}W=B\sqrt{\la \hat{N}_{\textrm{tot}} \ra}-C\la \hat{N}_{\textrm{tot}} \ra,\end{align}
where $B$ and $C$ are constants. Using $B$ and $C$ as free parameters, we fit this relation to the data in Fig. \ref{fig:epsart}, observing excellent agreement. We see then that the witness detects entanglement when the output is dominated by third-order correlations, but gradually fails as the even-order correlations grow.

To go beyond the weak-driving regime we use the input-output method with quantum pulses~\cite{Kiilerich2019}.
Following this, we derive a master equation to study the dynamics of the three resonator modes together with the three filtered output modes. Here, each propagating output wavepacket is treated as a single mode of an auxiliary fictitious resonator. 
We solve numerically the dynamics of 6 bosonic modes using QuTiP~\cite{Lambert2024}.
In~\cite{supplementary} we provide more details on this method and benchmark the numerical simulations against analytical results in the weak-driving regime. We also include Kerr nonlinearities, which we measure using a calibrated measurement of the cavity Kerr shifts.

We can directly compare the simulated entanglement witness value against that observed experimentally by setting the temporal profile of the propagating mode in the simulations to match the antialiasing filter of the digitizers and sweeping over the same range of drive strengths, $g$.
Since $g$ is not an experimentally accessible parameter, we estimate it by fitting the lowest order nonzero correlators $\langle \hat{N_i} \rangle$, in simulations to those observed experimentally, and find that $g$ is within 0.057 and 0.181~MHz~\cite{supplementary}. 
The numerically calculated witness values are shown in Fig.~\ref{fig:epsart} as the solid black line. Although the simulated value of $W$ quantitatively disagrees by a multiplicative factor of 1.9, the qualitative form is reproduced well and is in agreement with the analytical scaling predictions. Although we can improve the quantitative agreement by allowing additional free parameters, such as the pump detuning, to vary in the simulations, we felt it better to show the level of agreement with as many parameters as possible fixed at their independently measured and calibrated values. Given the experimental and theoretical complexity of the system, we find the agreement to be quite good.

\begin{figure}
    \centering
    \includegraphics[width=0.48\textwidth]{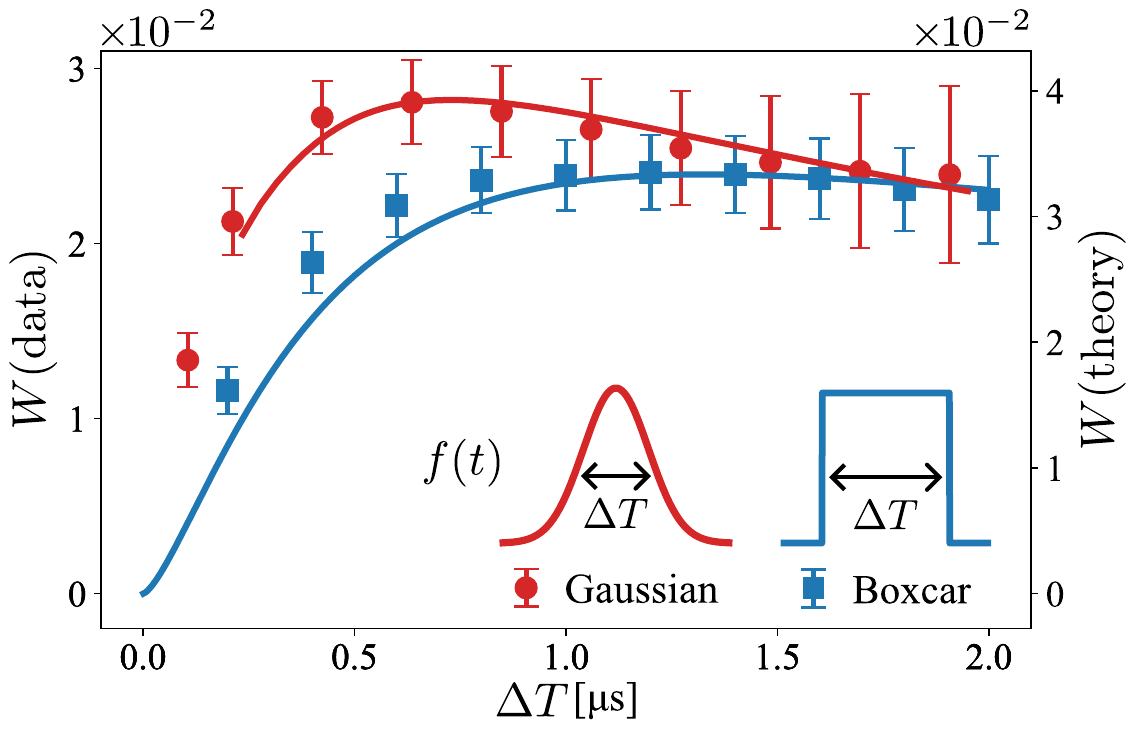}
    \caption{Entanglement witness, $W$, as a function of width, $\Delta T$, of the temporal mode function for Gaussian (red) and boxcar (blue) mode functions. Markers are experimental data, solid lines are theoretical predictions. The width of the Gaussian mode function is defined as the full-width-half-maximum value, while the boxcar width is defined by the edges of the boxcar. $W$ grows rapidly for small $\Delta T$, until it reaches a maximum. The maximum values are achieved when $\Delta T$ is significantly larger than the cavity lifetimes ($ 0.12-0.25 \, \mu$s).  We see that larger values of  $W$ are obtained with the Gaussian mode function, in agreement with previous works~\cite{ChalmersFilters, Lu2021}.} 
    \label{fig:WitnessVsModeFunc}
\end{figure}
To explore the nontrivial dependence of $W$ on the multimodal structure of the output state~\cite{Khanahmadi_Lund_Mølmer_Johansson_2023}, we test two types of temporal mode function. To do this experimentally, we first oversample the data with a large sampling frequency, and then digitally filter with the mode function of our choice. In the continuous pumping regime, discretization is implemented as a convolution of the measured quadrature voltages with the chosen temporal mode function~\cite{supplementary}. 
We use a boxcar as a simple reference function and compare it to a Gaussian function, which was shown in~\cite{ChalmersFilters} to reveal stronger nonclassical properties when studying the output state of a Kerr parametric oscillator.

The entanglement witness value as a function of temporal mode widths for both boxcar and Gaussian functions is shown in Fig.~\ref{fig:WitnessVsModeFunc}. For both functions, the witness grows to a maximum when the width is several times the average cavity lifetime of 0.17 $\mu$s, then drops off for wider functions. Similar to the parametric oscillator in~\cite{ChalmersFilters}, the observed entanglement witness is larger for Gaussian modes and peaks at a width around 5 times the average cavity lifetime. Although $W$ is not an entanglement monotone, and therefore does not necessarily measure the strength of the entanglement, we can infer that the relative magnitude of the tripartite correlations in our propagating mode reaches an optimum value at a finite width of temporal modes. This can be understood as the mode with the highest probability of capturing the emission of all three photons in a triplet into the transmission line.

The Gaussian entanglement generated by two-photon SPDC has been a key enabling technology for many experimental achievements in quantum information. The unique entanglement structure generated from non-Gaussian processes offers to expand these capabilities and enable novel applications and experiments in this domain. Among its nonclassical behaviours, the non-Gaussian tripartite entanglement of nondegenerate 3P-SPDC heralds its potential importance for quantum metrology~\cite{Fisher_3Photon}, probing the fundamentals of entanglement~\cite{NonG_Steering, From3PSPDC_toCats}, and even studying models of quantum gravity~\cite{ QuantumGravity_Sabin}. 
Integrating this device with superconducting qubits has been proposed to enable swapping of the non-Gaussian entanglement between bosonic oscillator modes and qubit modes~\cite{EntanglementSwapping_Sabin}. Furthermore, marginal improvements to the temperature of the cavity could enable the observation of the metrological advantage predicted in~\cite{Fisher_3Photon}. 
Our results open new avenues for exploring quantum states of light involving non-Gaussian entanglement in a continuous-variable system by leveraging the unique characteristics of 3P-SPDC in microwave quantum optics.

We thank Nicolai Friis, Marcus Huber, and Carlos Sabin for important discussions on the proper choice of witness.  We thank Joe Aumentado and the Advanced Microwave Photonics Group (686.05) at the National Institute of Standards and Technology (Boulder) for providing the shot noise tunnel junction device used for calibration. We further thank Joe Aumentado and Maxime Malnou at NIST for useful discussions on performing broadband calibrations. CMW, BJF, AS, IN, CWSC, and DD acknowledge the Canada First Research Excellence Fund (CFREF); NSERC of
Canada; the Canadian Foundation for Innovation; the Ontario Ministry of Research and Innovation; Defense Research and Development Canada; Industry Canada; and the Innovation for Defence Excellence and Security (IDEaS) program from the Canadian Department of National Defence (DND) for financial support.
FN is supported in part by: the Japan Science and Technology Agency (JST) [via the CREST
Quantum Frontiers program Grant No. JPMJCR24I2,
the Quantum Leap Flagship Program (Q-LEAP), and
the Moonshot R\&D Grant Number JPMJMS2061], and
the Office of Naval Research (ONR) Global (via Grant
No. N62909-23-1-2074).

\bibliography{main}

\end{document}


\preprint{APS/123-QED}
\bibliographystyle{apsrev4-2}
\title{Supplementary Material for ``Observation of Genuine Tripartite Non-Gaussian Entanglement from a Superconducting Three-Photon Spontaneous Parametric Down-Conversion Source"}

\maketitle

\onecolumngrid
\tableofcontents

\clearpage 

\onecolumngrid 
\renewcommand{\arraystretch}{1.5} 

\section{Cryogenic Setup}

\begin{figure}[h!]
    \centering
    \includegraphics[width=0.7\linewidth]{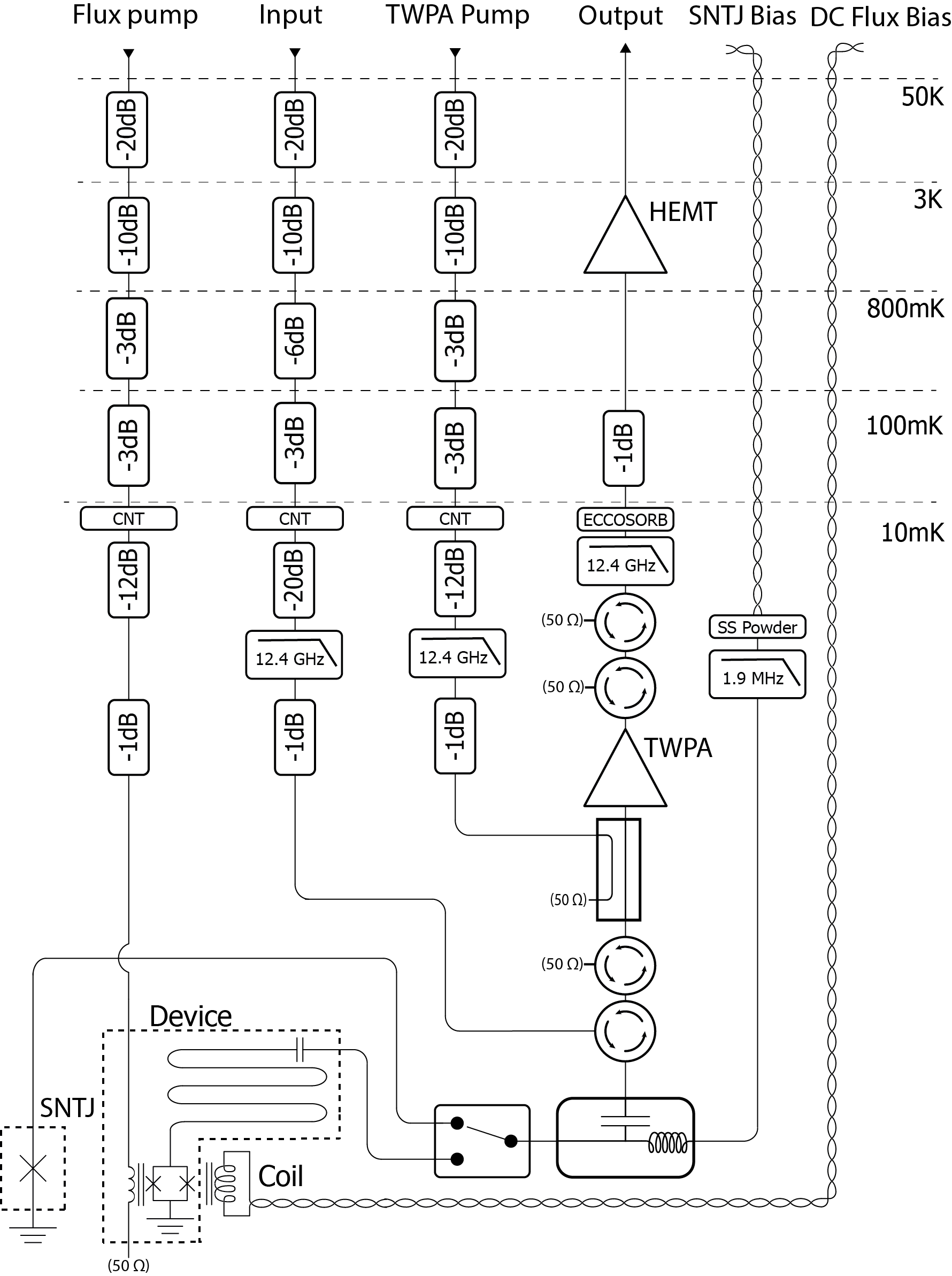}
    \caption{Diagram of cryogenic microwave network used to generate and measure propagating entangled photon triplets.}
    \label{fig:FridgeDiagram}
\end{figure}

The cryogenic setup used in this experiment is shown in figure \ref{fig:FridgeDiagram}. A high frequency tone is sent down the flux pump line to drive the SPDC process inside the device. Entangled photon triplets propagate out of the cavity through the TWPA and are measured at room temperature with three RF digitizers at the output line. The input line is used for characterizing the device with standard vector network analyzer (VNA) measurements. The DC flux bias is used to tune the static flux through the SQUID. The SNTJ bias line varies the voltage across the SNTJ to produce the shot noise curves used for gain and noise temperature calibration. The SNTJ is thermalized to the same plate as the device to enable measurement of the device temperature. The output line from the SNTJ to the RF switch is measured at room temperature before cool down to be identical to that from the device to the RF switch to ensure accurate gain calibration.

\section{Device}

\subsection{Cavity parameters}
We provide the frequencies, quality factors, decay rates, and lifetimes of the three cavity modes used in this experiment in table~\ref{table:Cavity_Params}.
\begin{table}[ht]
    \begin{center}
        \begin{tabular}{|c||c|c|c|c|c|} 
         \hline
        Frequency (GHz)  & $Q_{\text{ext}}$ & $Q_{\text{tot}}$ & $\gamma_{\text{ext}}$ (MHz) & $\gamma_{\text{tot}}$ (MHz) & $\tau$ ($\mu$s)\\ [0.5ex]  
          \hline\hline
        5.560 & 9205 & 8701 & 3.795 & 4.015 & 0.25\\
         \hline
        6.831 & 5500 & 5102 & 7.804 & 8.413 & 0.12\\
        \hline
        7.902 & 10593 & 7498 & 4.687 & 6.622 & 0.15\\
        \hline
        \end{tabular}
    \caption{Quality factors and cavity lifetimes, and decay rates of cavity modes.}
    \label{table:Cavity_Params}
    \end{center}
\end{table}

\subsection{Kerr matrix}
Here, we provide the matrix of Kerr coefficients for the three cavity modes used in our device in table \ref{table:Kerr_Matrix}. To measure the Kerr coefficients, we drive the cavity on resonance through the coupling capacitor and measure the power scattered off of the cavity using the calibrated system gain described in section \ref{Section:Calibration}. The drive populates the cavity modes with a known photon number and we simultaneously measure the induced shift in its resonance frequencies using a weak vector network analyzer probe tone. A symmetric Kerr matrix is required for the resulting terms in the Hamiltonian to be Hermitian. The asymmetry in the measured Kerr matrix is likely a manifestation of systematic errors in our experiment. The calculated Kerr terms are dependent on both the calibrated gain measurements and the resonator linewidths obtained from fitting the frequency responses of the modes when probed by a VNA. Since these frequency responses are measured against background ripples in our measurement transmission lines, we presume that there is some error in the fitted linewidths in addition to uncertainty in the calibrated gains. To maintain Hermiticity, the Kerr matrix used in our numerical simulations in section \ref{section:numerical_methods} is symmetrized, $K_\text{sym}=(K+K^T)/2$.
\begin{table}[ht]
    \begin{center}
        \begin{tabular}{|c| |c| c |c|} 
         \hline
        Pump$\backslash$Probe & 5.56 GHz& 6.83 GHz & 7.90 GHz \\ [0.5ex] 
          \hline\hline
        5.56 GHz& 36.4 & 70.1 & 136.0 \\
         \hline
        6.83 GHz& 45.4 & 28.7 & 107.4 \\
        \hline
        7.90 GHz& 139.5 & 172.6 & 160.3 \\
        \hline
        \end{tabular}
        \caption{A matrix of Kerr and cross-Kerr shifts all in units of KHz. The frequency on the left column is for the driving tone. The frequency on the top row is the mode being analyzed.}
        \label{table:Kerr_Matrix}
    \end{center}
\end{table}

\begin{figure}[h]
    \centering
    \includegraphics[width=0.8\linewidth]{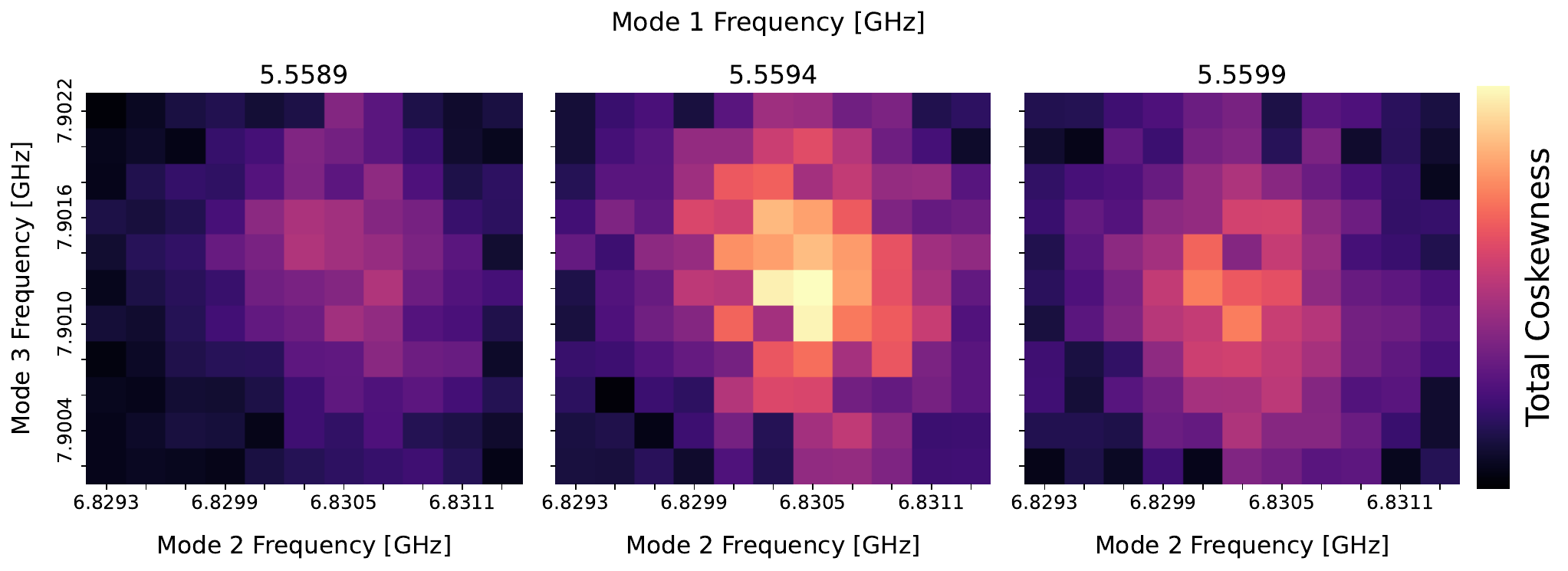}
    \caption{Three mode quadrature coskewness as a function of measurement frequency of each mode for a given pump power. The pump frequency is always set to the sum of the three mode frequencies. The optimal pump and probe conditions are taken to be at the point of maximum coskewness between all three modes.}
    \label{fig:Coskewness}
\end{figure}

\subsection{Tuning for optimal pump and detection frequencies}
We suppose that the optimal pump and detection frequencies for observing entanglement will also generate the largest third order correlations between the three modes in our device. We sweep the detection frequencies for the three modes while fixing $\omega_p=\omega_1+\omega_2+\omega_3$ and observe the room temperature correlations between all three modes as shown in figure \ref{fig:Coskewness}. We can then select good detection frequencies for observing entanglement without performing a full analysis of the quantum correlations. The heuristic for third order correlation that we measure here is simply the sum of absolute coskewnesses between all three modes for all possible combinations of quadratures I,Q. We perform this sweep at each pump power to track the shifting mode frequencies.

\subsection{Calibrating 3P SPDC coupling rate}
\begin{figure}[h]
    \centering
    \includegraphics[width=0.45\linewidth]{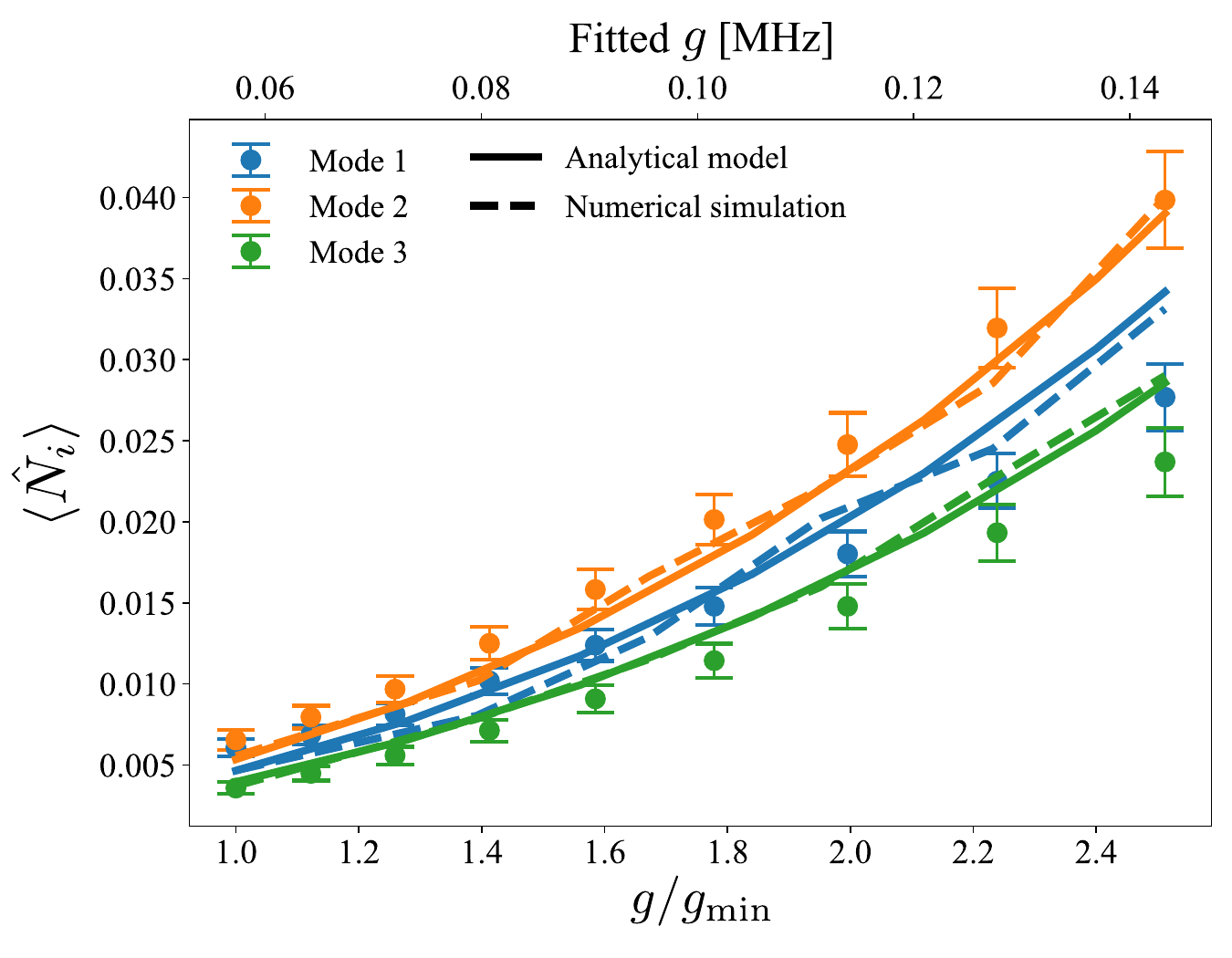}
    \caption{Numerical fitting of photon counts fitted to calibrate the 3P-SPDC coupling rate. The experimental value of $g/\gmin$ is obtained by taking the square root of the room temperature pump power and dividing by the minimum drive strength ($\gmin$) used. The top axis is the best estimate of the true value of $g$ found by the fit using the analytical model (solid line). The numerical simulation (dashed line) shows strong agreement with both the experimental data and the analytical model. Note that the fit is restricted to the range of data in the weak-driving regime, where the perturbative model remains valid ($g\ll \gamma_{\text{tot}}$). See sections \ref{section:perturbation_thry}, and \ref{section:numerical_methods} for details on analytical and numerical models.}
    \label{fig:MatchingPhotons}
\end{figure}

We do not have direct experimental access to the coupling rate, $g$, in the 3P-SPDC Hamiltonian. However, we know that $g \propto |\alpha_{p}| \propto \sqrt{P_{p}}$, where $P_p$ is the room temperature power of the tone sent down the pump line which we can convert to $g$ in normalized units by dividing by the minimum drive strength used in our measurement sweep, $\gmin$. Then, we can use a numerical fit to estimate the true values of $g$. We fit to the lowest order operator that we study, the output photon counts, using an analytic relation between the cavity photon population and the output photon population as described in section \ref{section:two-time_Correlations}. This allows us to perform an optimization without requiring the full quantum trajectory simulation used to simulate the higher order correlators in the output field as described in section \ref{section:numerical_methods}. As shown by the fit of the output photon population in Fig. \ref{fig:MatchingPhotons}, we estimate  $\gmin = 0.057$ MHz, giving the full parameter range considered in the main text to be from 0.057 to 0.181~MHz. 

\section{Measurements}
\subsection{Quantum description of measured fields}
We use the standard quantum mechanical formalism of amplification and linear field measurements to relate the room temperature voltage measurements taken at the RF digitizers to the quantum signal in the output field of the cavity. We model these combined processes as measuring the following non-Hermitian operator
\begin{equation}\label{eq:S_operator}
    \begin{split}
        \hat{S} &= \sqrt{\frac{G}{2}}\left(\hat{A} + \hd \right) \\ 
        & = \frac{1}{2} \left(\hat{X} + i\hat{P}\right),
    \end{split}
\end{equation}
where $\hat{h}$ is the bosonic operator for a noise mode including contributions from both the linear amplifiers and the mixer used in the digitizers. $\hat{X}$ and $\hat{P}$ are both Hermitian operators representing the two quadratures of the measured signal. In reality, the digitizer takes voltage measurements of the root-mean-squared in-phase ($I$) and quadrature ($Q$) components of the microwave signal where:
\begin{equation}
    \tilde{V}(t) = \frac{I(t) + iQ(t)}{\sqrt{2}}.
\end{equation}
The voltage quadratures can be quantized to the field quadratures via the transformation:
\begin{equation}
    \begin{split}
        I(t) & \rightarrow \frac{\sqrt{\hbar \omega Z_{0}}}{2}\hat{X}(t) \\
        Q(t) &\rightarrow \frac{\sqrt{\hbar \omega Z_{0}}}{2}\hat{P}(t),
    \end{split}
\end{equation}
where $Z_{0} = 50 \,\Omega$ is the characteristic impedance, and $\omega$ is the mode frequency.

\subsection{Discretization of continuum modes}
As mentioned in the main text, we use the standard technique to map continuum mode operators to discrete mode operators by defining a temporal mode function $f(t)$ such that the discrete mode, $\hat{A}_i$ is related to the continuum mode in the output field, $\hat{a}_{\text{out},i}$ as
\begin{align}
    \hat{A}_i  =\int \text{d}t\, f(t)\hat{a}_{\text{out},i}(t),
\end{align} 
where $f(t)$ is any real or complex function that is normalized as $\int_{0}^{\infty} {\rm d}t \, \vert f(t) \vert^2 = 1$ to preserve the bosonic commutation rules, and is equivalent to a wave packet for the propagating field.

Experimentally, we implement the temporal mode function in two ways: by using the built-in filtering of the RF digitizers, or by oversampling and digitally filtering in post-processing. In the first case, we do not perform any explicit filtering techniques because the RF digitizers have built-in anti-aliasing filters which are smooth and have a sharp cut-off in the frequency domain. However, to compare our numerical time domain simulations (see section \ref{section:numerical_methods}) to our experimental results, we need to translate the filter into an explicit temporal mode function.

\begin{figure}
    \centering
    \includegraphics[width=0.75\linewidth]{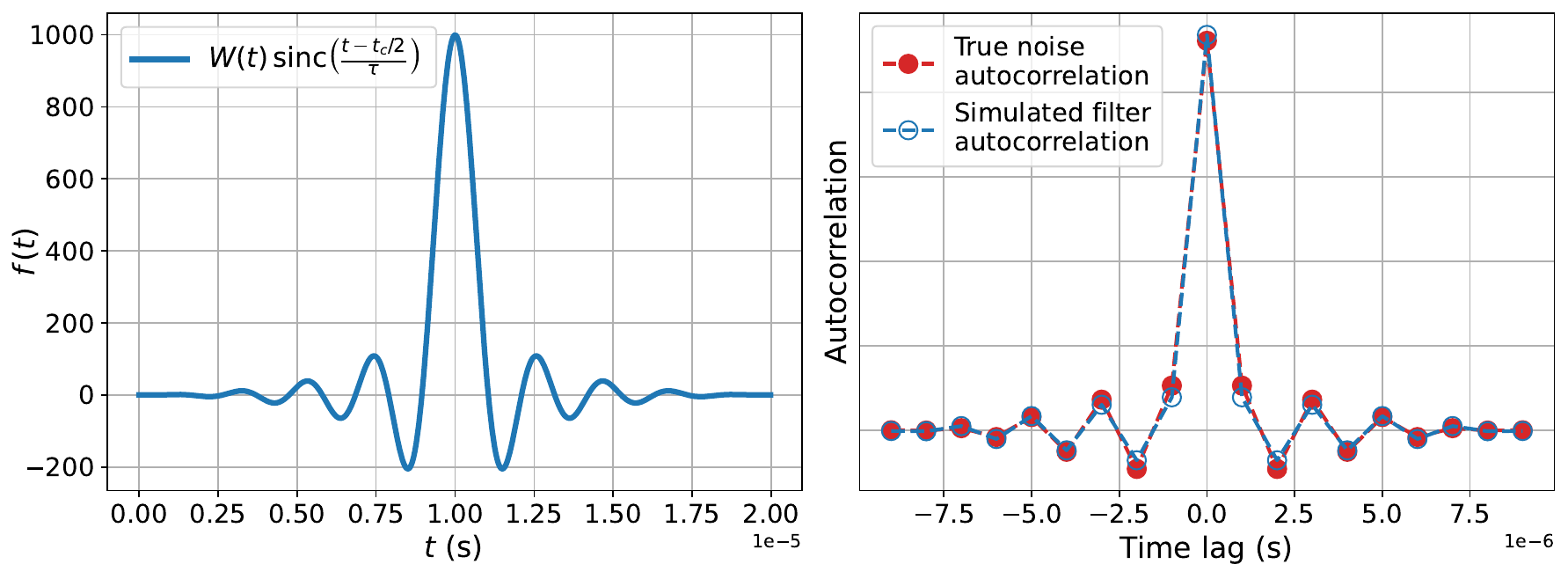}
    \caption{Estimating the effective temporal mode function of the built-in anti-aliasing filters in the RF digitizer. (left) Continuous temporal mode function where $W(t)$ is the window function specified in eq.~\eqref{eq:filter}. (right) Comparison of discrete autocorrelation function of filtered white noise with the autocorrelation function of our estimate of the time domain filter which shows strong agreement. The precision in our estimate is limited by the spacing between time lags which is given by $dt = 1/f_s$.}
    \label{fig:sinc_filtering}
\end{figure}

The anti-aliasing filter is well-approximated by a windowed sinc function in the time domain \cite{DSP_Oppenheim}. For the principal measurements using a sampling frequency of $f_s = 1$ MHz, we find that the temporal mode function is to a good approximation given by
\begin{equation}\label{eq:filter}
    f(t)= \sqrt{f_{s}}\, R(0,t_c)\,H(t_c)\,\text{sinc}\left(\frac{t-t_c/2}{\tau}\right),
\end{equation}
where $R(0,t_c) = \Theta(t)-\Theta(t-t_c)$ truncates the function outside $t=0$ and $t=t_c$ ($\Theta(t)$ is the Heaviside step function), and $H(T_c) = \text{sin}^{2}\left(\frac{\pi t}{t_c}\right)$ is the Hann window centered at $t=t_c/2$, and $\text{sinc}(x) \equiv \text{sin}(\pi x)/(\pi x)$. We confirm that this mode function with $t_c = 20 \;\mu \text{s}$, and $\tau = 1.05 \; \mu \text{s}$ matches the true time domain filter using the fact that the autocorrelation of filtered white noise is proportional to the autocorrelation of the mode filter function up to a constant factor (given by the variance of the noise) as shown in Fig. \ref{fig:sinc_filtering}. We also numerically confirm that $\int_{-\infty}^{\infty}|f(t)|^2 = 1$. 

In the second case, we use the fact that if we sample the signal with a frequency much larger than the decay rate of the cavity (i.e. $f_s \gg \gamma_{\text{ext}}/2\pi$) then we can treat the measured data as a full digital reconstruction of the continuous signal \cite{Eichler_Thesis, LinearDetection2}, and therefore we have the freedom to filter the signal arbitrarily in post-processing. This is done by convolving the measured quadratures with the desired temporal mode function
\begin{equation}
    \begin{split}
        X &= \sum_{\tau=0}^{N_{\text{coeff}}}X_{t}f_{t+\tau} \\
        P &= \sum_{\tau=0}^{N_{\text{coeff}}}P_{t}f_{t+\tau},
    \end{split}
\end{equation} 
where $f_{t}$ is the discretized mode function. It is clear from this that the resolution of the filter function is limited by the sampling frequency, and that each data point now requires $N_{\text{coeff}}$ times more measurements. We use this method to explore the impact of varying the width of different temporal mode functions (boxcar and Gaussian) on $W$ as shown in figure 4 in the main text.

Although the oversampling provides more freedom over which mode function we can use, it is very demanding because the number of data points for each measurement is multiplied by the number of coefficients in the mode function, which is typically of order 10-100. Furthermore, we observe only a minor increase in the entanglement witness value when using an optimized post-processing mode function compared to the mode function defined by the built-in filtering. Therefore, we chose to use the built-in filters with a 1 MHz sampling frequency for the sweeping measurement of $W$ as function of $\langle\hat{N}_{\rm tot} \rangle$ in figure 3 of the main text. 

\subsubsection{Using different mode functions for each mode}
\begin{figure}[h]
    \centering
    \includegraphics[width=0.42\linewidth]{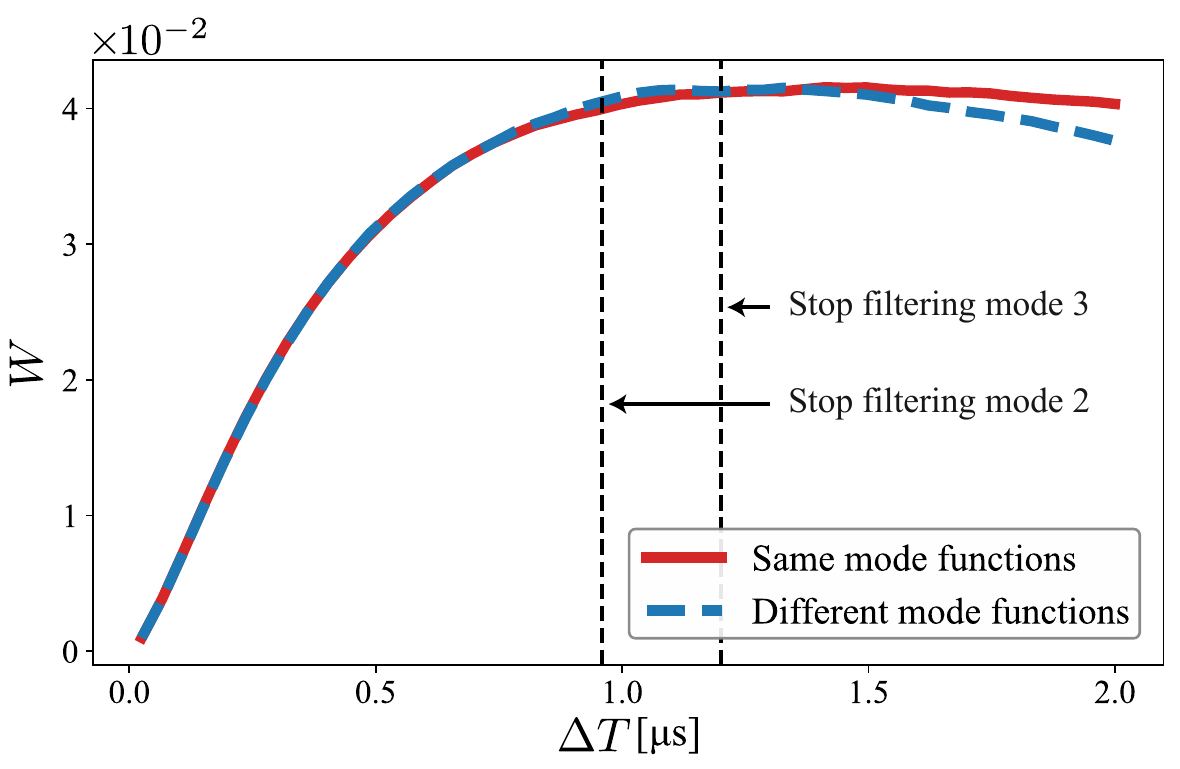}
    \caption{Simulation of entanglement witness as a function of boxcar temporal mode width where each mode has a different width (dashed blue) versus the same mode width (red). In the case of different mode width, each mode is filtered until a maximum width of 8 times its lifetime as indicated by the vertical dashed lines. The measured entanglement witness decreases when we begin using different mode widths for each mode relative to when we use the same width for all three.}
    \label{fig:diff_mode_functions}
\end{figure}

In the most general case, we can define a unique temporal mode function to discretize each mode. 
To investigate the significance of using different mode functions, we compare the effect of changing the width of each mode function using the numerical methods explained in section~\ref{section:numerical_methods}. The plot in Fig.~\ref{fig:diff_mode_functions}, shows that the entanglement witness drops off faster when using different temporal mode functions for each mode. Although not a comprehensive test, this suggests that there is no obvious advantage to using different mode functions for each mode.

\subsection{Converting measurements to expectations of moments}\label{Section:Correlators}
The general approach to obtaining the operator moments of the cavity output is to measure the field both when the 3P-SPDC drive is on, and when it's off, then perform a subtraction that moves the reference frame of our measurements to the output of the cavity. This is generally possible because the noise added by the amplification chain is Gaussian and delta correlated, $\braket{\hat{h}}=0$ and $\braket{\hat{h}\hat{\xi}} = \braket{\hd\hat{\xi}}=0$ for any other mode $\hat{\xi}$. To prevent drift in the amplification chain noise over long time scales from invalidating the assumption of delta correlation, we interleave the ON and OFF measurements in time every 1 second. Since the device is known to be at a nonzero temperature (see section \ref{Section:Calibration}), we also account for the contribution of the weak thermal state being emitted by the cavity even when the drive is OFF in our calculations of the moments of the field. The expectation of the number operator for the thermal state is given by the Bose-Einstein distribution 
\begin{equation}\label{eq:Bose-Einstein}
    \braket{\hat{N}_{T,i}} \equiv N_{T,i} = \frac{1}{e^{\hbar\omega_i/k_{\text{B}}T}-1},
\end{equation}
where $\omega_i$ is the frequency of mode $i$, and $T$ is the device temperature which we can measure directly with the calibration in section \ref{Section:Calibration}. This gives us the moments of the persistent thermal contribution to the output field that we measure and which we account for when calculating the moments of the output field. In the following subsections, we derive the relationship between the room temperature measurements of $\hat{S}_i$ and the three moments used to calculate our entanglement witness. 

\subsubsection{$\braket{\hat{N}_{i}}$}

\begin{equation}\label{eq:2ndOrderMoment}
    \begin{split}
        \braket{\Sd_{i}\hat{S}_{i}}_{\rm on}
        &= G_{i}\left[ \braket{\hat{N}_{i}} + \braket{\hat{N}_{N,i}} \right] \\
        \braket{\Sd_{i}\hat{S}_{i}}_{\rm off} &= G_{i}\left[ N_{T,i} + \braket{\hat{N}_{N,i}} \right] \\
        \\ 
        \Rightarrow \braket{\hat{N}_{i}} &= \frac{\braket{\Sd_{i}\hat{S}_{i}}_{\rm on} - \braket{\Sd_{i}\hat{S}_{i}}_{\rm off}}{G_{i}} + N_{T,i}.
    \end{split}
\end{equation}

\subsubsection{$\braket{\hat{N}_{i}\hat{N}_{j}}$}

\begin{equation}
    \begin{split}
            \braket{\Sd_{i}\hat{S}_{i}\Sd_{j}\hat{S}_{j}}_{\rm on} & = G_{i}G_{j}\left[\braket{\hat{N}_{i}\hat{N}_{j}} + \braket{\hat{N}_{i}}\braket{\hat{N}_{N,j}} + \braket{\hat{N}_{j}}\braket{\hat{N}_{N,i}} + \braket{\hat{N}_{N,i}}\braket{\hat{N}_{N,j}}\right] \\
            \braket{\Sd_{i}\hat{S}_{i}\Sd_{j}\hat{S}_{j}}_{\rm off} & = G_{i}G_{j}\left[N_{T,i}N_{T,j} + N_{T,i}\braket{\hat{N}_{N,j}} + N_{T,j}\braket{\hat{N}_{N,i}} + \braket{\hat{N}_{N,i}}\braket{\hat{N}_{N,j}}\right] \\ 
            \\ 
            \Rightarrow \braket{\hat{N}_{i}\hat{N}_{j}} &= \frac{\braket{\Sd_{i}\hat{S}_{i}\Sd_{j}\hat{S}_{j}}_{\rm on}   \braket{\Sd_{i}\hat{S}_{i}\Sd_{j}\hat{S}_{j}}_{\rm off}}{G_{i}G_{j}} + N_{T,i}N_{T,j}   \\
            & - \braket{\hat{N}_{N,j}}\left(\braket{\hat{N}_{i}} - N_{T,i} \right) - \braket{\hat{N}_{N,i}}\left(\braket{\hat{N}_{j}} - N_{T,j} \right), 
    \end{split}
\end{equation}\vspace{10px}

where $\braket{\hat{N}_{i}}$ is known from eq.~\eqref{eq:2ndOrderMoment}, and $\braket{\hat{N}_{N,i}} = \braket{\Sd_{i}\hat{S}_{i}}_{\rm off}/G_{i} - N_{T,i}$. 

\subsubsection{$\braket{\hat{A}_{1}\hat{A}_{2}\hat{A}_{3}}$}
\begin{equation}
    \begin{split} \braket{\hat{S}_{1}\hat{S}_{2}\hat{S}_{3}}
        & = \sqrt{G_{1}G_{2}G_{3}}\braket{\hat{A}_{1}\hat{A}_{2}\hat{A}_{3}} \\
        \\
        \Rightarrow \braket{\hat{A}_{1}\hat{A}_{2}\hat{A}_{3}} & = \frac{\braket{\hat{S}_{1}\hat{S}_{2}\hat{S}_{3}}}{\sqrt{G_{1}G_{2}G_{3}}}.
    \end{split}
\end{equation}

\subsection{Measured moments}
Here we include an example of the measured moments where we observed a maximum value of the entanglement witness (as highlighted in figure 3 of the main text) in table \ref{table:WTable_BE}.

\begin{table*}[h!]
    \begin{center}
        \begin{tabular}{|c| |c| c |c| c|} 
         \hline
         Mode Indices  & $\abs{\langle\hat{A}_1\hat{A}_2\hat{A}_3\rangle} \times10^{-2}$ & $\langle\hat{N}_i\rangle\times10^{-2}$ & $\langle\hat{N}_j\hat{N}_k\rangle\times10^{-3}$ & $W\times10^{-2}$\\ [0.5ex] 
         \hline \hline
        $i=1$, $j=2$, $k=3$&  $ 2.680 \pm 0.099$ & $ 1.018 \pm 0.082 $ & $ 1.48\pm0.22 $ & $ 1.554\pm0.103 $  \\
         \hline
        $i=2$, $j=3$, $k=1$&  & $ 1.250\pm0.100$ & $ 0.80\pm0.21 $&  \\
        \hline
        $i=3$, $j=1$, $k=2$&  & $ 0.713\pm0.067$ & $ 2.49\pm0.45$&  \\
        \hline
    \end{tabular}
    \caption{Measured moments of the output field under optimal pump tuning and a $1 \,\text{MHz}$ sampling frequency obtained using the best estimate of the system gain explained in section \ref{Section:Calibration}. The reported value for the entanglement witness of $W=\left(1.554\pm0.103\right)\times 10^{-2}$ shows that we observe $ W>0$ with over 15 standard deviations of confidence.}
    \label{table:WTable_BE}
    \end{center}
\end{table*}

\subsection{Certifying that the observed state is non-Gaussian}

Theoretically, the photon triplet states are generated via a nonquadratic Hamiltonian and therefore they are non-Gaussian states. Here, we verify that this is the case for the states observed in our experiment. We do this by considering a Gaussian state with the measured first- and second-order moments, and showing that it is not compatible with the measured value of $\la\hat{a}_1\hat{a}_2\hat{a}_3 \ra$. In particular, we will show that, for a Gaussian state, 
\begin{equation}\label{GaussianRequirement}
    \la \hat{a}_1\hat{a}_2\hat{a}_3 \ra = \langle \hat{a}_1 \rangle \langle \hat{a}_2\hat{a}_3 \rangle + \langle \hat{a}_2 \rangle \langle \hat{a}_1\hat{a}_3 \rangle + \langle \hat{a}_3\rangle \langle \hat{a}_1\hat{a}_2 \rangle - 2 \langle \hat{a}_1 \rangle \langle \hat{a}_2 \rangle \langle \hat{a}_3 \rangle.
\end{equation}

We take as given that this type of third-order correlator is zero for any mean-zero Gaussian state. We then use the fact that any Gaussian state with finite mean can be obtained by a displacement of a zero-mean Gaussian state. Using the three-mode displacement operator,
\begin{equation}
    \begin{split}
           \hat{D}(\alpha_1, \alpha_2,\alpha_3) &= \text{exp}(\alpha_1 \hat{a}_1^{\dg} - \alpha_1^{\ast} \hat{a}_1) \text{exp}(\alpha_2 \hat{a}_2^{\dg} - \alpha_2^{\ast} \hat{a}_2)\text{exp}(\alpha_3 \hat{a}_3^{\dg} - \alpha_3^{\ast} \hat{a}_3),
    \end{split}
\end{equation}
we can write the finite-mean state $\hat{\rho}$ in terms of a mean-zero state $\hat{\rho}_0$  as $\hat{\rho} = \hat{D}(\alpha_1, \alpha_2,\alpha_3) \hat{\rho}_0 \hat{D}^{\dg}(\alpha_1, \alpha_2,\alpha_3)$. The expectation value $\langle \hat{a}_1\hat{a}_2\hat{a}_3 \rangle$ on the finite-mean Gaussian state can be written as
\begin{equation}
    \begin{split}
       \langle \hat{a}_1\hat{a}_2\hat{a}_3 \rangle & \equiv \Tr [ \hat{a}_1\hat{a}_2\hat{a}_3 \hat{\rho}] \\ 
       & =   \Tr[ \hat{a}_1\hat{a}_2\hat{a}_3 \hat{D}(\alpha_1, \alpha_2,\alpha_3) \hat{\rho}_0 \hat{D}^{\dg}(\alpha_1, \alpha_2,\alpha_3)] \\
       & = \Tr[\hat{D}^{\dg}(\alpha_1, \alpha_2,\alpha_3)\hat{a}_1\hat{a}_2\hat{a}_3\hat{D}(\alpha_1, \alpha_2,\alpha_3) \hat{\rho}_0] \\
       & = \Tr[(\hat{a}_1 + \alpha_1) (\hat{a}_2 + \alpha_2) (\hat{a}_3 + \alpha_3) \hat{\rho}_0] \\
        & = \langle \hat{a}_1\hat{a}_2\hat{a}_3 \rangle_0 + \langle \hat{a}_1 \alpha_2 \alpha_3 \rangle_0 + \langle \hat{a}_2 \alpha_1 \alpha_3 \rangle_0 + \langle \hat{a}_3 \alpha_1 \alpha_2 \rangle_0  \\
        & \; \; \; \; + \langle \hat{a}_1\hat{a}_2 \alpha_3 \rangle_0 + \langle \hat{a}_1\hat{a}_3 \alpha_2 \rangle_0 + \langle \hat{a}_2\hat{a}_3 \alpha_1 \rangle_0 + \langle \alpha_1 \alpha_2 \alpha_3 \rangle_0,
    \end{split}
\end{equation}
where we have defined $\langle\hat{O}\rangle_0 $ as the expectation of the operator $\hat{O}$ on the mean-zero Gaussian state. For a mean-zero Gaussian state, $\langle \hat{a}_1\hat{a}_2\hat{a}_3 \rangle_0 = \langle \hat{a}_1\rangle_0 = \langle \hat{a}_2 \rangle_0 = \langle \hat{a}_3  \rangle_0=0$, and so
\begin{equation}
    \langle \hat{a}_1\hat{a}_2\hat{a}_3 \rangle = \alpha_3\langle \hat{a}_1\hat{a}_2 \rangle_0 + \alpha_2\langle \hat{a}_1\hat{a}_3  \rangle_0 + \alpha_1\langle \hat{a}_2\hat{a}_3  \rangle_0 + \alpha_1 \alpha_2 \alpha_3. 
\end{equation}
Finally, we can use the same arguments as above to do the reverse transformation for the second-order moments to produce
\begin{equation}
        \langle \hat{a}_1\hat{a}_2\hat{a}_3 \rangle = \alpha_3 \langle \hat{a}_1\hat{a}_2 \rangle + \alpha_2\langle  \hat{a}_1\hat{a}_3 \rangle + \alpha_1 \langle \hat{a}_2\hat{a}_3 \rangle - 2 \alpha_1 \alpha_2 \alpha_3.
\end{equation}

To confirm whether our experimentally observed state is indeed non-Gaussian, we consider the moments for the measurements corresponding to the maximum value of $W$ observed in the text. The first- and second-order moments are given in table~\ref{table:NG_Test}.
\begin{table*}[h!]
    \begin{center}
        \begin{tabular}{|c||c|c|} 
         \hline
         Mode Indices  & $\langle \hat{A}_i \rangle \times 10^{-5}$ & $\langle \hat{A}_i\hat{A}_j \rangle\times 10^{-3}$  \\ [0.5ex] 
         \hline \hline
        $i=1$, $j=2$&  $ 4.35\pm6.5 $ & $ 6.54\pm0.24 $    \\
         \hline
        $i=2$, $j=3$& $-3.20\pm4.3$ & $ 4.57\pm0.20$    \\
        \hline
        $i=3$, $j=1$& $-4.97\pm3.6$ & $ 4.59\pm0.18$    \\
        \hline
    \end{tabular}
    \caption{Measured first, second, and third moments of the output field.}
    \label{table:NG_Test}
    \end{center}
\end{table*}

Comparing magnitudes, we have:
\begin{equation}
    \begin{split}
        \abs{\langle \hat{A}_1\hat{A}_2\hat{A}_3 \rangle} &= (2.68 \pm 0.099)\times10^{-2} \\ 
        \abs{\langle \hat{A}_1 \rangle \langle \hat{A}_2\hat{A}_3 \rangle + \langle \hat{A}_2 \rangle \langle \hat{A}_1\hat{A}_3 \rangle + \langle \hat{A}_3\rangle \langle \hat{A}_1\hat{A}_2 \rangle - 2 \langle \hat{A}_1 \rangle \langle \hat{A}_2 \rangle \langle \hat{A}_3 \rangle} &= (2.73 \pm4.28)\times10^{-7},
    \end{split}
\end{equation}
where the measured third-order correlator is taken from the main text. These values are distinct with 27$\sigma$ of confidence. That is, the observed value of  $|\langle \hat{A}_1\hat{A}_2\hat{A}_3 \rangle|$ is strongly incompatible with the Gaussian construction, certifying that our observed state is non-Gaussian.

\section{Absolute System Calibration with SNTJ\label{Section:Calibration}}

Absolute system calibration is performed using a shot noise tunnel junction (SNTJ) as a calibrated noise source. This is a standard method to calibrate a measurement chain consisting of well-behaved linear amplifiers \cite{SNTJScience, SNTJExample, SNTJReview, SNTJ10mk, SNTJRef1}. However, the broadband characteristic of an SNTJ makes it challenging to calibrate a near-quantum-limited parametric amplifier such as a TWPA due to both signal and idler mixing and gain compression. In this section, we describe how we can account for these effects to make both a best estimate, and an upper bound estimate on the absolute system gain.

\subsection{Standard amplifier calibration}
The total noise power emitted from the SNTJ is a combination of voltage-dependent shot noise and constant Johnson-Nyquist noise. The room-temperature power observed over a narrow bandwidth is given by: 

\begin{equation}
    \begin{split}\label{eq:SNTJ_RT}
         P = G \, \text{BW} \, k_{\text{B}} \biggl\{ T_{N} +  \frac{1}{2} \biggl[ & \left( \frac{eV + hf}{2k_{\text{B}}}\right) \text{coth}\left( \frac{eV + hf}{2k_{\text{B}}T} \right) + \\ 
         & \left( \frac{eV - hf}{2k_{\text{B}}}\right) \text{coth}\left( \frac{eV - hf}{2k_{\text{B}}T} \right) \biggr] \biggr\},
    \end{split}    
\end{equation}
where $G$ is the total gain in the amplifier chain, BW is the measurement bandwidth, $T_N$ is the noise temperature of the linear amplifier, and $T$ is the physical temperature of the SNTJ. By fitting the data with eq.~\eqref{eq:SNTJ_RT}, we can extract the absolute gain, noise temperature, and physical temperature of the full output line shown in Fig~\ref{fig:FridgeDiagram}.

\subsection{Parametric amplifier calibration}\label{section:Para-amp_Calibration}
 Parametric amplifiers mix together signal and idler modes in order to amplify the signal. Since the SNTJ produces broadband noise, the idler mode is often as hot as the signal mode during calibration, which must be accounted for in our fitting procedure. Starting with the standard model of phase-insensitive parametric amplification we have:
\begin{equation}
    a^{s}_{\text{out}} = \sqrt{\sGain} \, a^{s}_{\text{in}} + \sqrt{\sGain-1}\, a^{i \dagger}_{\text{in}},
\end{equation}
where $a^s$ and $a^i$ are the signal and idler modes, respectively, and $G^{s\rightarrow s}$ is the signal-to-signal gain. It is common to approximate $\sqrt{\sGain} \approx \sqrt{\sGain-1}$ but we keep the full expression here because the TWPA gain can be on the order of 10. Accounting for added noise in the amplifier and shifting to units of power, we have
\begin{equation}
    P^{s}_{\rm out} = \sGain \left(P^{s}_{\rm in} + P^{s}_{N} \right) + (\sGain-1) P^{i}_{\rm in} 
\end{equation}
where $P^{s}_N$ is the excess noise power added by the amplifier at the signal frequency. In reality, there is always loss both in the pre-amplifier components, and inside the amplifier. We can absorb the loss into the gain as:
\begin{equation}
    \begin{split}
       & \seffG = \sGain \, \eta^{s} \\
       & \ieffG = (\sGain-1) \, \eta^{i},  
    \end{split}
\end{equation}
where $\tilde{G}$ is the effective gain, and $\eta^{s},\eta^{i} \leq 1$ are the signal and amplifier transmission coefficients, respectively. The total output power of the amplifier is then:
\begin{equation}\label{eq:IdlerGain}
    P^{s}_{\rm out} = \tilde{G}^{s\rightarrow s} \left(P^{s}_{\rm in} + P^{s}_{N} + \frac{\tilde{G}^{i\rightarrow s}}{\tilde{G}^{s\rightarrow s}}P^{i}_{\rm in} \right).
\end{equation}

It is clear from equation \eqref{eq:IdlerGain} that the idler mode can contribute significantly to the output power if $P^{i}_{\text{in}} \sim P^{s}_{\text{in}}$ which is the case for a broadband noise source like the SNTJ. The gain will be strictly overestimated if the idler mode input is not accounted for, which means that we can find an upper bound estimate on the gain by assuming $P^{i}_{\text{in}} = 0$ which we use to quantify the maximum systematic error in the entanglement witness (see section \ref{sec:SystematicError}). 

\subsubsection{Estimating $\tilde{G}^{i\rightarrow s}/\tilde{G}^{s\rightarrow s}$}
In principle, an exact calibration can be done given the frequencies of the signal and idler modes and the effective gain ratio, $\tilde{G}^{i\rightarrow s}/\tilde{G}^{s\rightarrow s}$. For a 4-wave mixing TWPA as used in this experiment, the signal and idler frequencies are determined by the pump via $2\omega_p = \omega_s + \omega_i$.
However, we pump the TWPA nondegenerately and so the signal and idler modes are separated by many GHz. Since $\tilde{G}^{i\rightarrow s}/\tilde{G}^{s\rightarrow s} \approx \eta^{i}/\eta^{i}$ is determined by frequency-dependent losses in both the TWPA and the pre-amplifier stages, which are not directly accessible experimentally, we cannot generally measure it. A solution is to use numerical fitting to estimate the ratio. From equation \eqref{eq:SNTJ_RT}, the quantum noise floor relative to the voltage induced shot-noise is frequency dependent and thus the mixture of the signal and idler noise given by the ratio $\tilde{G}^{i\rightarrow s}/\tilde{G}^{s\rightarrow s}$, can be included as a fitting parameter independent of $\tilde{G}^{s\rightarrow s}$. With a reasonable initial guess, the model should be able to fit for $\tilde{G}^{i\rightarrow s}/\tilde{G}^{s\rightarrow s}$ because it is uniquely determined by the onset of shot noise dominating over quantum noise with respect to applied voltage since the signal and idler frequencies are known. 

The fitted ratios for the signal and idler modes as determined by a 8.326 GHz pump at the TWPA are shown in table \ref{table:Gain_Ratio}. The contribution of the 11.09 GHz idler mode to the 5.56 GHz signal mode is small due to the much higher insertion loss at 11 GHz of the microwave components between the device output and the TWPA (see fig. \ref{fig:FridgeDiagram}). Similarly, the idler modes for the 6.83 and 7.90 GHz signal contribute about half as much as the signal mode. 

\begin{table}[ht]
    \begin{center}
        \begin{tabular}{|c|c|c|} 
         \hline
        Signal frequency (GHz) & Idler frequency (GHz) &  $\tilde{G}^{i\rightarrow s}/\tilde{G}^{s\rightarrow s}$ \\ [0.5ex] 
          \hline\hline
        5.56 & 11.09 &  0.228 $\pm$ 0.003 \\
         \hline
        6.83 & 9.82 &  0.478 $\pm$ 0.033 \\
        \hline
        7.90 & 8.75 & 0.570 $\pm$ 0.240 \\
        \hline
        \end{tabular}
        \caption{Fitted values of $\tilde{G}^{i\rightarrow s}/\tilde{G}^{s\rightarrow s}$ for each mode.}
        \label{table:Gain_Ratio}
    \end{center}
\end{table}

\subsection{Calibration with gain compression}
A separate challenge of calibrating a Josephson junction-based TWPA with a broadband noise source is the relatively low saturation power. Again, following the techniques in \cite{SNTJReview}, we can model the power output of an amplifier as a function of the input power according to
\begin{equation}
    P_{\rm out}(P_{\rm in}) = G(P_{\rm in})\, P_{\rm in},
\end{equation}
where the gain is a function of the form
\begin{equation}
    G(P_{\rm in}) = g_{\rm ss}\, \lambda(P_{\rm in}),
\end{equation}
with $g_{\rm ss}$ being the small signal gain and $\lambda(P_{\rm in}) \leq 1$ being the gain compression profile of the amplifier. We typically want to operate in the region where $\lambda = 1$ for a linear response in the amplifier. However, the broadband power from the SNTJ when biased into the shot noise dominated regime is relatively high and causes compression in the otherwise linear power with respect to bias voltage. 

When accounting for gain compression, equation \eqref{eq:IdlerGain} becomes

\begin{equation}\label{eq:Gain_Compression}
    P^{s}_{\rm out} = \tilde{g}^{s\rightarrow s} \,\lambda(P_{\rm in})\left(P^{s}_{\rm in} + P^{s}_{N} + \frac{\tilde{g}^{i\rightarrow s}}{\tilde{g}^{s\rightarrow s}}P^{i}_{\rm in} \right),
\end{equation}
where we have assumed that the signal-to-signal and idler-to-signal gain compression profiles are identical (which we measured to be true to within a few percent for our system). If we have $\lambda(P_{\rm in})$, we can simply divide the measured output power to get $P^{s}_{\rm out} /\lambda(P_{\rm in})$, and then perform a numerical fit as we would with eq. \eqref{eq:IdlerGain}.

\subsubsection{Measuring the gain compression profile}\label{section:CompressionProfile}
\begin{figure}[h]
    \centering
    \includegraphics[width=1\textwidth]{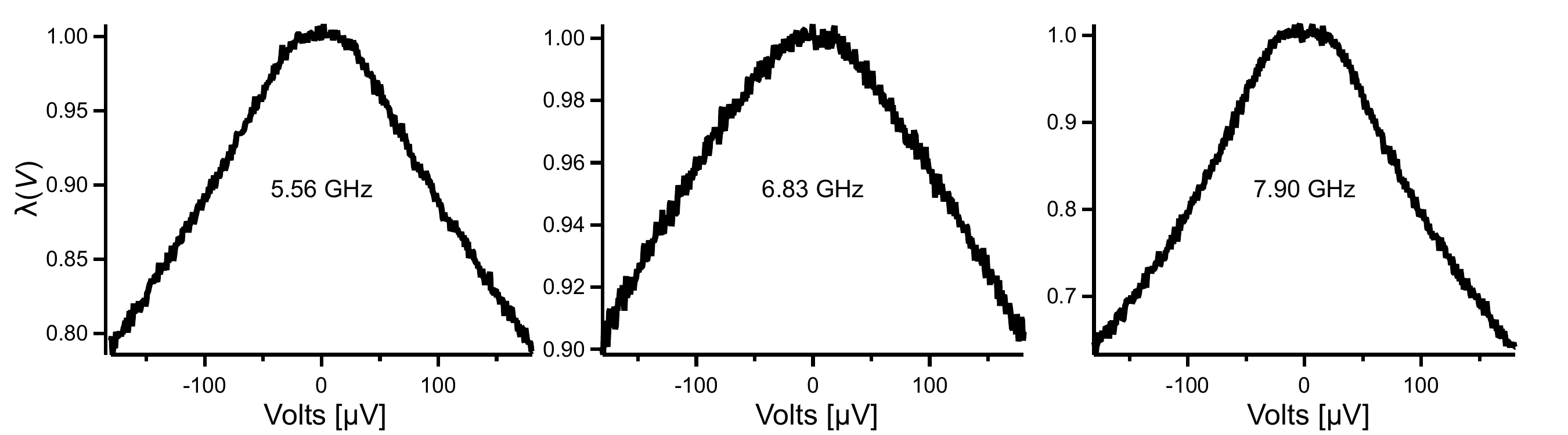}
    \caption{Gain compression profiles of the output line at the three mode frequencies. Note that there is a frequency-dependent compression likely due to the complex mixing processes inside the TWPA.}
    \label{fig:Compression_Profile}
\end{figure}

If we send a coherent probe with constant power $P^{\rm probe}_{\rm in}$ through the amplifier chain while the SNTJ is on, we observe the total power output in the measurement bandwidth:
\begin{equation}
    P_{\rm out} = G(V)\, (P^{\rm probe}_{\rm in} + P^{\rm SNTJ}_{\rm in}(V)),
\end{equation}
where we write $G(V)$ because in the weak probe limit we can assume that the compression is entirely governed by the voltage at the SNTJ. To decouple the probe from the SNTJ, we can just subtract the output powers:
\begin{equation}
\begin{split}
     P^{\rm probe}_{\rm out}  & = G(V)\,  [P_{\rm out} - P^{\rm SNTJ}_{\rm out} (V)] \\
     & = G(V)\, P^{\rm probe}_{\rm in} ,      
\end{split}  
\end{equation}
which can be normalized by its maximum value: $P^{\rm probe,max}_{\rm out} = g_{\rm ss}\, P^{\rm probe}_{\rm in} $ ,  giving the gain compression profile:
\begin{equation}
\begin{split}
        P_{\rm norm}(V)  &= \frac{P^{\text{probe}}_{\rm out}}{P^{\text{probe, max}}_{\text{out}} }  = \frac{G(V)\, P^{\text{probe}}_{\rm in}}{g_{\text{ss}}\, P^{\text{probe}}_{\text{in}}} \\
        & = \lambda(V),
\end{split}
\end{equation}
which we can then use to re-scale the output power and extract the small signal gain. The measured gain compression profiles are shown in Fig.~\ref{fig:Compression_Profile}. 

\subsection{Calibrated gain values and fits}
\begin{figure}
    \centering
    \includegraphics[width=1\linewidth]{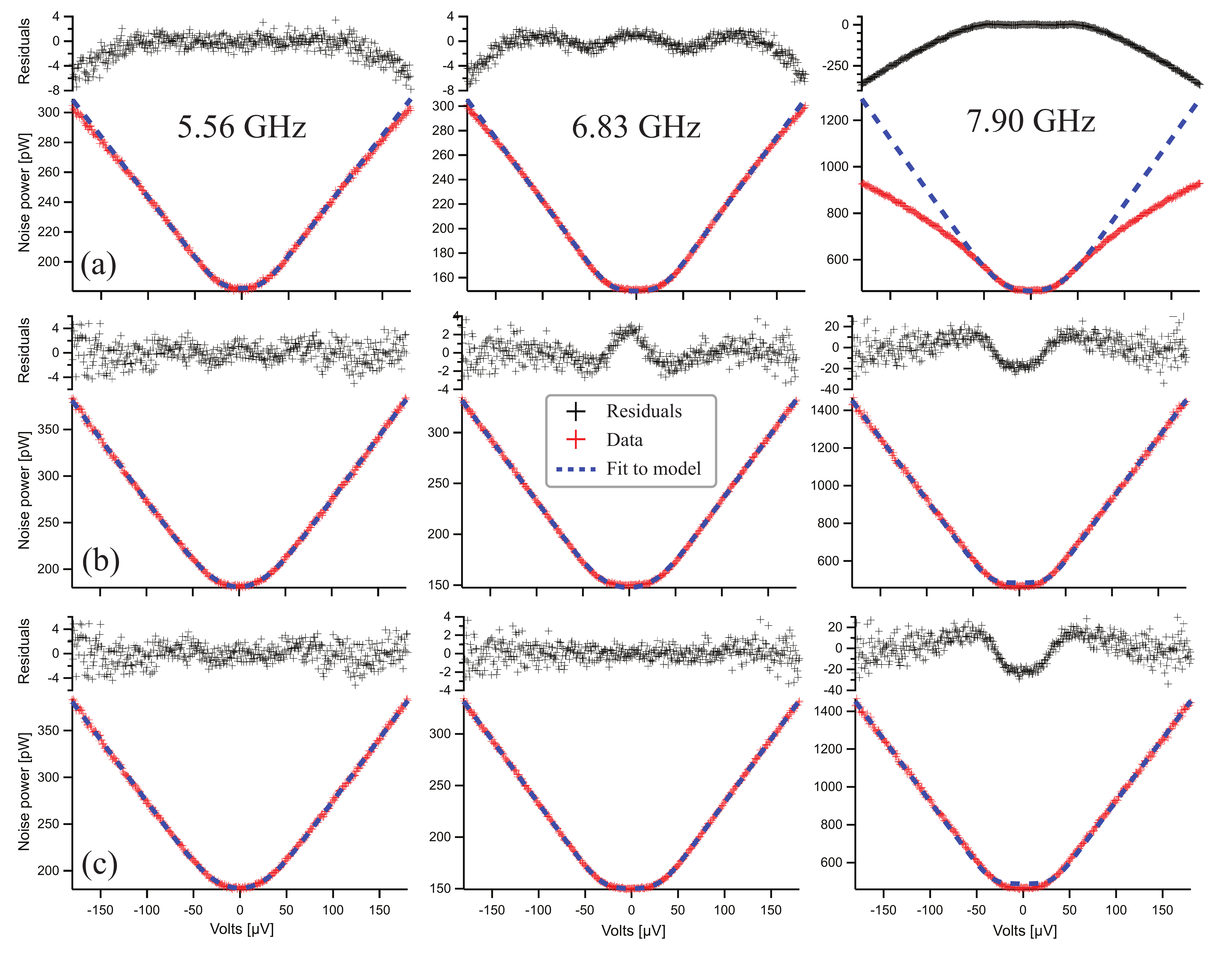}
    \caption{Fitted output power from the STNJ at three mode frequencies using three different models for parametric amplification of broadband noise. Residuals for each fit are shown above in black. (a) Best fit without correction for gain compression or idler-to-signal mixing. The model overshoots the data in the linear shot noise region due to the gain compression, especially for the 7.90 GHz data. (b) Best fit with correction for gain compression, but not for idler-to-signal mixing. This model fits the linear shot noise region well due to the compression correction, but the peaks in the residuals at the center of the 5.56 GHz and 6.83 GHz data suggest that the model cannot properly fit the quantum noise region due to the idler-to-signal mixing. (c) Best fit with correction for both gain compression and idler-to-signal mixing. The model now fits the full range of data. The small dip in the residuals for the 7.90 GHz fit is indicative of more complex parametric processes near the dispersive feature of the TWPA that cannot be fully accounted for by the model even with the corrections. The fitted gains for all three models are shown in table \ref{table:Calibrated_Gains}.}
    \label{fig:Full_Calibration}
\end{figure}

To demonstrate the importance of accounting for idler-to-signal mixing and gain compression in the gain calibration, we plot the best fits for three cases in Fig.~\ref{fig:Full_Calibration}. It's clear from the fits and the trend of the residuals in Fig.~\ref{fig:Full_Calibration} a) that the gain compression causes curvature in the shot noise which cannot be accounted for by the compression free model. The fits significantly improve when we account for gain compression as shown in Fig.~\ref{fig:Full_Calibration} b), especially for the 7.90 GHz data. However, the transition from the quantum to shot noise region is still not fully accounted for as demonstrated by the peak in the residuals at 0 bias for the 5.56 and 6.83 GHz data. We attribute this to idler-to-signal mixing with a higher frequency and thus higher quantum noise idler mode. This is largely corrected when we include the idler mode as shown by the very good fits for the 5.56 and 6.83 GHz data shown in Fig.~\ref{fig:Full_Calibration} c). The model has a slight issue fitting the center of the 7.90 GHz which is likely due to some more complex nonlinear processes near the dispersive feature of the TWPA, but the linear shot noise region is very well fitted by the model, so the gain value should be accurate.

Table \ref{table:Calibrated_Gains} shows the best fit gain values in all three cases. Note that the column corresponding to the data corrected for gain compression but without idler contribution has the largest gain values, as we expected, because it is the upper bound estimate of the gain, and thus corresponds to a systematic lower bound estimate for the entanglement witness as described in section \ref{sec:SystematicError}). The third column is our best possible estimate for the true gain, and thus corresponds to the ``best estimate" of the witness.

\begin{table}[ht]
    \begin{center}
        \begin{tabular}{|c| |c| c |c|} 
         \hline
        Frequency (GHz) & Uncorrected & Corrected without idler & Corrected with idler \\ [0.5ex] 
          \hline\hline
        5.56 GHz& 70.03 & 72.23 & 71.31 \\
         \hline
        6.83 GHz& 71.09 & 71.86 & 70.16 \\
        \hline
        7.90 GHz& 78.48 & 79.13 & 77.31 \\
        \hline
        \end{tabular}
        \caption{Calibrated gain values in dBm when using the raw data versus the data corrected for gain compression with and without including the contribution of idler to signal mixing in the amplifier. We use the values in the ``corrected with idler" column to obtain the entanglement witness in the main text. We use the values in the ``corrected without idler" column to obtain a lower bound estimate of the entanglement witness due to systematic uncertainty (see section \ref{sec:SystematicError}).}
        \label{table:Calibrated_Gains}
    \end{center}
\end{table}

\subsubsection{Physical device temperature}
\begin{figure}
    \centering
    \includegraphics[width=0.5\linewidth]{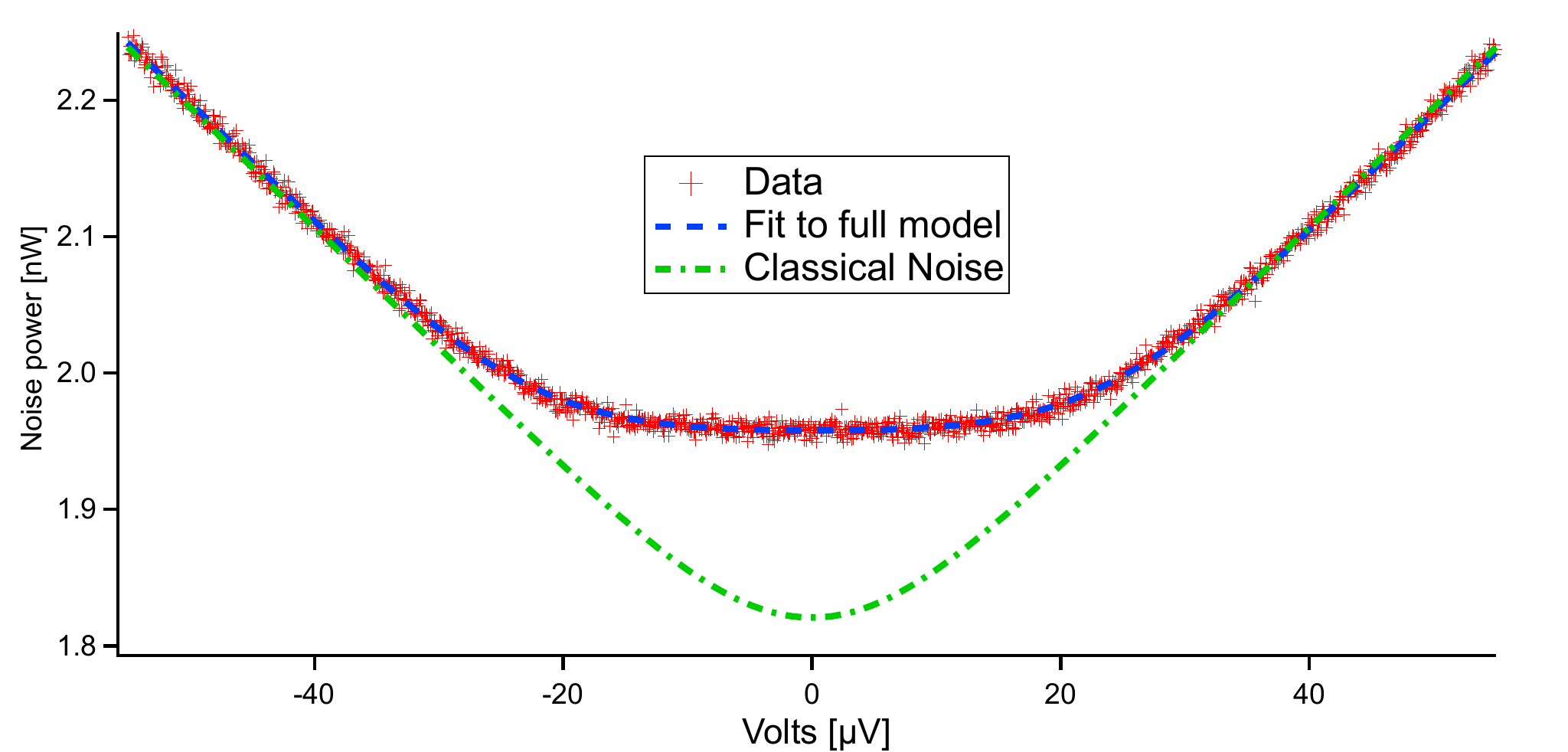}
    \caption{Output power from the SNTJ (red) and best fit (dashed blue) with faster sampling frequency and smaller voltage range at the SNTJ to best capture the shot noise to quantum noise transition which is required to calibrate the physical temperature of the device.}
    \label{fig:Physical_temp}
\end{figure}

The gain calibration data was taken in the middle of the measurement sweep at a 1 MHz sampling frequency to most accurately measure the gain during the sweep. However, it does not provide the most accurate estimate of the physical temperature of the device because the SNTJ is not in thermal equilibrium with the device immediately after switching the RF switch, and many data points are needed near the curvature point to reliably measure the physical temperature. Thus, we take a separate measurement with a higher sampling frequency (10 MHz) and a smaller voltage range ($\pm 50 \, \mu V$) after the measurement sweep is finished to calibrate the cavity temperature most accurately. Fig.~\ref{fig:Physical_temp} shows the fitted data at 5.56 GHz used to calibrated the physical temperature with a comparison to the classical noise curve. We measure the effective temperatures at 5.56, 6.83, and 7.90 GHz to be: 43.7$\pm0.7$, 41.1$\pm0.3$, and 45.5$\pm0.3$ mK, respectively. From the Bose-Einstein distribution in eq. \eqref{eq:Bose-Einstein} we can easily see that the thermal contribution at each frequency is relatively small, but not completely negligible, hence why we include it in our calculation of the correlators in section \ref{Section:Correlators}.

\section{Error analysis}
\subsection{Statistical error propagation of $W$}\label{section:SatisticalError}
To quantify the statistical error in the moments, we take the value computed for each chopped measurements as a sample of the expectation. We then treat each expectation value as a sampled random variable, $X$ with a mean, $\bar{X}$, and error given by its variance, $\sigma^2_X$. We can then propagate this error to $W$ with the assumption that $X$ satisfies $\sigma_X \ll \bar{X}$.  For readability, we substitute the following variables 
\begin{align*}
    X_i\equiv{\la \hat{N_i}\ra},\\
    Y_i\equiv{\la \hat{N_j}\hat{N_k}\ra},\\
    Z\equiv{\la \hat{A}_i\hat{A}_j\hat{A}_k\ra},
\end{align*}
With $i,j,k=1,2,3$ and $i\neq j\neq k\neq i$. Our witness is then measured as
\begin{equation}
    W=\bar{Z}-\sum_{i=1}^{3}\sqrt{\bar{X}_i}\sqrt{\bar{Y}_i},
\end{equation}
with variance
\begin{align}\label{piecewise_W_expression}
    \Var\left(W\right)=\Var\left(Z\right)-2\sum_{i=1}^{3}\Cov\left(Z,\XY\right)+\sum_{i,j=1}^{3}\Cov\left(\XY,\XYj\right).
\end{align}

As per the standard technique of error propagation, we expand the nonlinear function of $X_i$, $\sqrt{X_i}$, to a polynomial around $\bar{X_i}$ with the variable $\tilde{X_i}\equiv X_i-\bar{X_i}$ which we truncate to 2nd order

\begin{align}
    \sqrt{X}&\approx\sqrt{\bar{X}}+\frac{1}{2\sqrt{\bar{X}}}\tilde{X}-\frac{1}{8\sqrt{\bar{X}}^3}\tilde{X}^2,\\
\sqrt{Y}&\approx\sqrt{\bar{Y}}+\frac{1}{2\sqrt{\bar{Y}}}\tilde{Y}-\frac{1}{8\sqrt{\bar{Y}}^3}\tilde{Y}^2.
\end{align}
We can now treat each term in (\ref{piecewise_W_expression}) and express them using moments we directly measure. Keeping terms up degree 2 in $\tilde{X},\ \tilde{Y}$ and $\tilde{Z}$ we have

\begin{align}
\Cov\left(\XY,\XYj\right)\approx\sqrt{\bar{X_i}\bar{X_j}\bar{Y_i}\bar{Y_j}}&\left(\frac{\Cov(X_i,X_j)}{4\bar{X_i}\bar{X_j}}
+\frac{\Cov(X_i,Y_j)}{4\bar{X_i}\bar{Y_j}}
+\frac{\Cov(X_j,Y_i)}{4\bar{X_j}\bar{Y_i}}
+\frac{\Cov(Y_j,Y_j)}{4\bar{Y_j}\bar{Y_j}}\right),
\end{align}
and
\begin{align}
    \Cov\left(Z,\XY\right)\approx\frac{\sqrt{\bar{X}}}{2\sqrt{\bar{Y}}}\Cov(Z,Y)+\frac{\sqrt{\bar{Y}}}{2\sqrt{\bar{X}}}\Cov(Z,X).
\end{align}
Combining these, we find the variance of the witness value is\begin{align}
    \Var(W)=&\Var(Z)-\frac{\sqrt{\bar{X}}}{\sqrt{\bar{Y}}}\Cov\left(Z,Y\right)-\frac{\sqrt{\bar{Y}}}{\sqrt{\bar{X}}}\Cov\left(Z,X\right)\nonumber\\
& \sum_{i,j=1}^{3} \sqrt{\bar{X_i}\bar{X_j}\bar{Y_i}\bar{Y_j}}\left(\frac{\Cov(X_i,X_j)}{4\bar{X_i}\bar{X_j}}
+\frac{\Cov(X_i,Y_j)}{4\bar{X_i}\bar{Y_j}}
+\frac{\Cov(X_j,Y_i)}{4\bar{X_j}\bar{Y_i}}
+\frac{\Cov(Y_j,Y_j)}{4\bar{Y_j}\bar{Y_j}}\right).
\end{align}

\subsection{Systematic Error}\label{sec:SystematicError}
The principal sources of systematic error are the imprecision in absolute system calibration and the slow-varying gain drift in the amplifiers which we attribute to the HEMT. 

\begin{figure}
    \centering
    \includegraphics[width=0.5\linewidth]{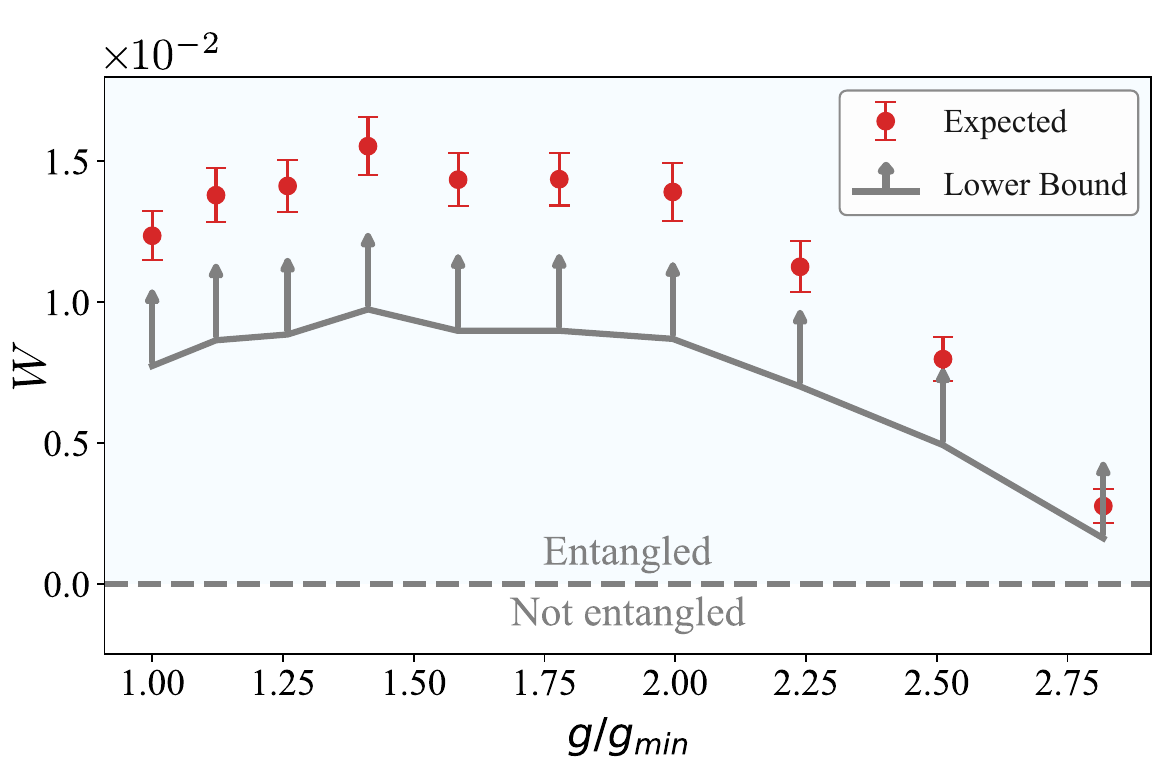}
    \caption{Lower bound on the estimate of the entanglement witness, $W$, as a function of three-photon drive strength, $g$, showing the maximum possible systematic error due to gain calibration. The data point where $W<0$ is not included.}
    \label{fig:W_LowerBound}
\end{figure}

We account for the imprecision in absolute system calibration by using a strict upper bound estimate on the gain to demonstrate that we observe entanglement in our system even if there is some error in our best estimate of the true gain. By looking at the calculations of the correlators in section \ref{Section:Correlators}, it's easy to see that:
\begin{align}\label{eq:WvsG}
    W\propto \frac{1}{\sqrt{G_1G_2G_3}},
\end{align}
up to some constant offset due to the thermal correction. This inverse square root proportionality means that any overestimation of the gain directly causes an underestimation of the entanglement witness. As such, we can use the fact that we measure a strict upper bound gain value for each mode using the gain calibration technique described in section \ref{section:Para-amp_Calibration} to set a lower bound on $W$. The systematic lower bound on $W$ relative to the reported value is shown on Fig.~\ref{fig:W_LowerBound}, which shows that we would observe entanglement ($W > 0$) even using the lower bound estimate.

To quantify the uncertainty due to gain drift, we apply a weak probe tone through the TWPA pump line at the three mode frequencies in the same way as described in section \ref{section:CompressionProfile} such that the output power is strictly dependent on the gain from the amplifier chain. We then record the output power every minute for about 24 hours to match the measurement period used to compute the entanglement witness. The relative uncertainty of the coherent power over these periods is the same as the relative uncertainty in gain used for the calculation of the entanglement witness: 
\begin{equation}
    \begin{split}
        \delta P_\text{probe}& \equiv \frac{\sqrt{\Var(P_\text{probe})}}{P_\text{probe}} \\ 
        & =\frac{\sqrt{\Var(G})} {{G}}\equiv\delta G.
    \end{split}
\end{equation}

We then obtain gain variances for each of the three modes and can perform an error propagation on these relative uncertainties. From eq.~\eqref{eq:WvsG}, the relative error in $G$ due to propagated long timescale gain uncertainty is:
\begin{align}
    \delta W =\frac{\delta G_2+\delta G_1+\delta G_3}{2}.
\end{align}

It is not impossible that the coherent power source also fluctuates adding its own contribution to the relative variance of the noise power. However, this can only lead to us overestimating the drift in gain which means we are at least placing an upper bound on this uncertainty. We find that over time bins equal to our measurement period, the relative uncertainty is: $\delta W=3.5\%$. 

The total uncertainty in $W$ due to both systematic error and statistical uncertainty is given by
\begin{equation}
    \Delta W = \sqrt{\text{Var}(W)} + W\cdot\delta W,
\end{equation}
where we take a linear sum of the two errors under the conservative assumption that there could be perfect correlation between drifts of different timescales.

\section{Steady-state perturbation theory}\label{section:perturbation_thry}

We study the dissipative dynamics described by the master equation
\begin{align}
    \frac{{\rm d}\hat{\rho}}{{\rm d}t} = \left( \mathcal{L}_0 + \mathcal{L}_1 \right)[\hat{\rho}],
\end{align}
with the dissipator
\begin{align}
    \mathcal{L}_0 [\hat{\rho}] &= \sum_j \frac{\gamma_j}{2} \left( 2 \hat{a}_j \hat{\rho} \hat{a}^\dagger_j - \hat{a}^\dagger_j\hat{a}_j \hat{\rho} - \hat{\rho} \hat{a}^\dagger_j\hat{a}_j \right),
\end{align}
which describes three-independent environments at zero temperature for the three bosonic modes $\hat{a}_i$, $i=1,2,3$. The steady state corresponding to this dissipative dynamics is $\hat{\rho}_{\rm ss} = \ket{000}\bra{000}$.
On the other hand, the superoperator $\mathcal{L}_1$ describes the unitary evolution under a nondegenerate three-photon drive
\begin{align}
    \mathcal{L}_1 [\hat{\rho}] &= -i g \left[ \hat{a}_1 \hat{a}_2 \hat{a}_3 + \hat{a}^\dagger_1 \hat{a}^\dagger_2 \hat{a}^\dagger_3,\hat{\rho} \right] .
\end{align}
Here, we restrict to the weak driving regime in which the three-photon drive strength $g$ is small compared to the total decay rate $\gamma_T = \sum_i \gamma_i$ of the three cavity modes, i.e., $g \ll \gamma_T$. In this regime, we can calculate perturbative corrections to the photon vacuum state due to the three-photon drive. In this analysis we ignore the presence of Kerr and cross-Kerr interaction terms as they are negligible in the few-photon regime.

We consider the following perturbative expansion for the steady-state density matrix
\begin{align}
    \hat{\rho}_{\rm ss} &= \hat{\rho}^{(0)}_{\rm ss} + \lambda \hat{\rho}^{(1)}_{\rm ss} + \lambda^2 \hat{\rho}^{(2)}_{\rm ss} + \mathcal{O}(\lambda^3),
\end{align}
where $\lambda \equiv g/\gamma_T$, i.e., the ratio between the three-photon drive strength to the total decay rate is the expansion parameter. 
Here $\hat{\rho}^{(0)}_{\rm ss} = \ket{000} \bra{000}$ corresponds to the steady-state of $\mathcal{L}_0$, i.e., $\mathcal{L}_0 \hat{\rho}^{(0)}_{\rm ss} = 0$. The different order corrections can be calculated by iteratively solving
\begin{align}
    \mathcal{L}_0 [\hat{\rho}_{\rm ss}^{(n)}] &= -\mathcal{L}_1 [\hat{\rho}_{\rm ss}^{(n-1)}] ,
\end{align}
for $n \geq 1$. 

\subsection{First-order correction}

To begin, we need to solve 
\begin{align}
    \mathcal{L}_0 [\hat{\rho}_{\rm ss}^{(1)}] &= ig \left( \ket{111} \bra{000} - \ket{000} \bra{111} \right) .
\end{align}
The above equation is satisfied by an ansatz of the form $\hat{\rho}_{\rm ss}^{(1)} \propto i(\ket{111}\bra{000} - \ket{000} \bra{111})$. From this the first-order correction immediately follows
\begin{align}
    \hat{\rho}^{(1)}_{\rm ss} = i \left( \frac{2g}{\gamma_T} \right) \left( \ket{000}\bra{111} - \ket{111} \bra{000} \right) .
\end{align}

\subsection{Second-order correction}

This time we need to solve
\begin{align}\label{eq:2nd}
    \mathcal{L}_0 [\hat{\rho}_{\rm ss}^{(2)}] &= \left( \frac{4g^2}{\gamma_T}  \right) \left[ \ket{000}\bra{000} - \ket{111}\bra{111} + \sqrt{2} \left( \ket{222} \bra{000} + {\rm h.c.} \right) \right].
\end{align}
It is straightforward to verify that an ansatz of the form   
\begin{align}
    \hat{\rho}^{(2)}_{\rm ss} &= \beta \ket{111}\bra{111} + \delta (\ket{222} \bra{000} + \ket{000}\bra{222}) + \epsilon \ket{011} \bra{011} + \theta \ket{101} \bra{101} + \psi \ket{110} \bra{110}  \nonumber\\
    &+ \kappa \ket{100} \bra{100} + \xi \ket{010} \bra{010} + \chi \ket{001} \bra{001} 
\end{align}
generates  all of the necessary terms on the right hand side of \eqref{eq:2nd} upon the action of $\mathcal{L}_0$. Solving for all of the individual terms, we obtain the expansion parameters
\begin{align}
    \beta &= \left( \frac{2g}{\gamma_T} \right)^2, \\
    \delta &= - \sqrt{2}  \left( \frac{2g}{\gamma_T} \right)^2, \\
    \epsilon &= \left( \frac{2g}{\gamma_T} \right)^2 \frac{\gamma_1}{\gamma_2 + \gamma_3}, \\
    \theta &= \left( \frac{2g}{\gamma_T} \right)^2 \frac{\gamma_2}{\gamma_1 + \gamma_3}, \\
    \psi &= \left( \frac{2g}{\gamma_T} \right)^2 \frac{\gamma_3}{\gamma_1 + \gamma_2}, \\
    \chi &= \left( \frac{2g}{\gamma_T} \right)^2 \frac{\gamma_1 \gamma_2}{\gamma_3} \left( \frac{1}{\gamma_1 + \gamma_3} + \frac{1}{\gamma_2 + \gamma_3}\right), \\
    \xi &= \left( \frac{2g}{\gamma_T} \right)^2 \frac{\gamma_1 \gamma_3}{\gamma_2} \left( \frac{1}{\gamma_1 + \gamma_2} + \frac{1}{\gamma_2 + \gamma_3}\right), \\
    \kappa &= \left( \frac{2g}{\gamma_T} \right)^2 \frac{\gamma_2 \gamma_3}{\gamma_1} \left( \frac{1}{\gamma_1 + \gamma_2} + \frac{1}{\gamma_1 + \gamma_3}\right) .
\end{align}

\subsection{Expectation values calculations}

Armed with the first- and second-order perturbative corrections to the steady-state, we calculate the different correlators contributing to the entanglement witness studied in the main text. The absolute value of the three-photon correlator is given by
\begin{align}
    \vert \langle \hat{a}_1 \hat{a}_2 \hat{a}_3 \rangle \vert &= \frac{2g}{\gamma_T}.
\end{align}
Only the first-order correction contributes to the above expression, while the second-order correction contributes to both, the mean number of photons 
$\langle \hat{n}_i \rangle = \langle \hat{a}^\dagger_i \hat{a}_i \rangle$
\begin{align}
    \langle \hat{n}_i \rangle &= \left( \frac{2g}{\gamma_T} \right)^2 \frac{\gamma_T}{\gamma_i} ,
\end{align}
and the number-number correlators $\langle \hat{n}_i \hat{n}_j \rangle = \langle \hat{a}^\dagger_i \hat{a}_i \hat{a}^\dagger_j \hat{a}_j \rangle$
\begin{align}
    \langle \hat{n}_i \hat{n}_j \rangle &= \left( \frac{2g}{\gamma_T} \right)^2 \left( \frac{\gamma_T}{\gamma_i + \gamma_j} \right) .
\end{align}
In the following section we show how to calculate the above correlators for the filtered output field. 

\section{Perturbative steady-state output field correlations}\label{section:output}

Correlators or expectation values of 
products of the filtered output field operators
$\hat{A}_i = \int_0^\infty {\rm d}t' \, f(t') \hat{a}_{{\rm out},i}(t')$ and their Hermitian conjugates result from the integration of multi-time output field correlations. 
For instance, the mean number of photons in mode $i$, $\langle \hat{N}_i \rangle = \langle \hat{A}^\dagger_i \hat{A}_i \rangle$ is given by the following relation
\begin{align}
    \langle \hat{A}^\dagger_i \hat{A}_i \rangle &= \int_0^\infty {\rm d}t'' \, \int_0^\infty {\rm d}t' \, f^*(t'') f(t') \langle \hat{a}^\dagger_{{\rm out},i}(t'') \hat{a}_{{\rm out},i}(t') \rangle_{\rm ss} .
\end{align}
For simplicity, we restrict to real-valued filter functions, i.e., $f^*(t) = f(t)$.
In turn, multi-time correlations can be calculated from the master equation governing the dynamics of the three-mode system by first, using the input-output relation to map output modes into cavity modes. For the case of an environment at zero-temperature, output multi-time correlators trivially map into cavity ones via the replacement $\hat{a}^{(\dagger)}_{{\rm out},i}(t) \to \sqrt{\gamma_{{\rm ext},i}} \, \hat{a}^{(\dagger)}_i(t)$, where $\gamma_{{\rm ext},i}$ is the external decay rate of mode $i$~\cite{Gardiner1985}. 
Here, we only consider steady-state correlations. 
As we detail in the following subsections, this means that the steady-state is taken as an initial condition to calculate the different order multi-time correlations. 
Once again, we consider our second order approximation to the steady-state and also treat the time evolution under the full Liouvillian in a perturbative fashion.

\subsection{Steady-state output field two-time correlations}\label{section:two-time_Correlations}

As seen above, the two-time correlations required to calculate the output field mean number of photons are 
$\langle \hat{a}^\dagger_{{\rm out},i}(t'') \hat{a}_{{\rm out},i}(t') \rangle_{\rm ss} = \gamma_{{\rm ext}, i } \langle \hat{a}^\dagger_i(t'') \hat{a}_i(t') \rangle_{\rm ss} $, with
$\gamma_{{\rm ext},i}$ the external decay rate of mode $i$.
Considering $t'' > t'$, the two-time cavity field correlator is given by~\cite{Gardiner1985}
\begin{align}\label{eq:two-time-propagator}
    \langle \hat{a}_i^\dagger(t'') \hat{a}_i(t') \rangle_{\rm ss} &= {\rm tr} \left[ \hat{a}_i^\dagger {\rm e}^{\mathcal{L} (t'' - t')} \hat{a}_i \hat{\rho}_{\rm ss} \right] ,
\end{align}
with $\mathcal L$, the total Liouvillian. In our case we can write
$\mathcal{L} = \mathcal{L}_0 + \mathcal{L}_1$ (see above), where $\mathcal{L}_1$ is treated as a perturbation. Up to first-order in $\mathcal{L}_1$, the propagator $V(t)\equiv\exp(\mathcal{L}t)$ can be approximated by~\cite{Villegas2016}
\begin{align}\label{eq:prop-expansion}
     V(t) &= V_0(t) + V_1(t) + \mathcal{O}(\lambda^2) \\
     &={\rm e}^{\mathcal{L}_0 t} + {\rm e}^{\mathcal{L}_0 t} \int_0^t {\rm d}t_1 \, 
    {\rm e}^{-\mathcal{L}_0 t_1} \mathcal{L}_1 {\rm e}^{\mathcal{L}_0 t_1}
     + \mathcal{O}(\lambda^2).
\end{align}
Notice that $V_0(t)$ corresponds to the ``free" propagator, that is, the propagator in the absence of a perturbation. We can use the above expression to solve \eqref{eq:two-time-propagator} for mode 1, i.e., $\hat{a}_i =\hat{a}_1$
\begin{align}
    \langle \hat{a}_1^\dagger(t'') \hat{a}_1(t') \rangle_{\rm ss} &= 
    {\rm tr} \left \{ \hat{a}_1^\dagger \left[ {\rm e}^{\mathcal{L}_0 (t''-t')} + {\rm e}^{\mathcal{L}_0 (t''-t')} \int_0^{(t''-t')} {\rm d}t_1 \, 
    {\rm e}^{-\mathcal{L}_0 t_1} \mathcal{L}_1 {\rm e}^{\mathcal{L}_0 t_1}
    \right] \left( \hat{a}_1 \hat{\rho}_{\rm ss} \right) \right \} \\
     &= 
    {\rm tr} \left[ \hat{a}_1^\dagger {\rm e}^{\mathcal{L}_0 (t''-t')} 
     \left( \hat{a}_1 \hat{\rho}_{\rm ss} \right) \right] + 
      {\rm tr} \left \{ \hat{a}_1^\dagger \left[  {\rm e}^{\mathcal{L}_0 (t''-t')} \int_0^{(t''-t')} {\rm d}t_1 \, 
    {\rm e}^{-\mathcal{L}_0 t_1} \mathcal{L}_1 {\rm e}^{\mathcal{L}_0 t_1}
    \right] \left( \hat{a}_1 \hat{\rho}_{\rm ss} \right) \right \} . \label{eq:two-time-expansion}
\end{align}
Looking at the first term on the right hand side of \eqref{eq:two-time-expansion}, a quick inspection of our second-order approximation to the steady-state suffices to realize that the terms $\beta \ket{111}\bra{111}$, $\theta \ket{101} \bra{101}$, $\psi \ket{110} \bra{110}$ and $\kappa \ket{100} \bra{100}$ give the first nonzero order ($\lambda^2$ )contribution to \eqref{eq:two-time-expansion}. 
Using the relation
\begin{align}\label{eq:liouvillian-relation1}
    \mathcal{L}_0\ket{0}_i \bra{n} &= -\frac{n\gamma_i}{2} \ket{0}_i\bra{n},
\end{align}
as well as
\begin{align}\label{eq:liouvillian-relation2}
    {\rm e}^{\mathcal{L}_0 t} \ket{1}_i\bra{1} &= {\rm e}^{-\gamma_i t} \ket{1}_i\bra{1} + (1 - {\rm e}^{-\gamma_i t}) \ket{0}_i\bra{0} ,
\end{align}
it is straightforward to calculate
\begin{align}
    {\rm tr} \left[ \hat{a}_1^\dagger {\rm e}^{\mathcal{L}_0 (t''-t')} 
     \left( \hat{a}_1 \hat{\rho}_{\rm ss} \right) \right] &= {\rm e}^{-\gamma_1 (t''-t')/ 2} (\beta + \theta + \psi + \kappa) .
\end{align}

In the second term on the right hand side of~\eqref{eq:two-time-expansion} the only correction of order $\lambda^2$ originates from the term $-i (2g/\gamma_T) \ket{111}\bra{000}$.
Similarly to the previous case, it is straightforward to show that
\begin{align}
    {\rm tr} \left \{ \hat{a}_1^\dagger \left[  {\rm e}^{\mathcal{L}_0 (t''-t')} \int_0^{(t''-t')} {\rm d}t_1 \, 
    {\rm e}^{-\mathcal{L}_0 t_1} \mathcal{L}_1 {\rm e}^{\mathcal{L}_0 t_1}
    \right] \left( \hat{a}_1 \hat{\rho}_{\rm ss} \right) \right \} &= \left( \frac{2g}{\gamma_T} \right)^2 \frac{\gamma_T}{\gamma_2 + \gamma_3 - \gamma_1}\left[ {\rm e}^{- \gamma_1 (t''-t')/2} - {\rm e}^{-(\gamma_2 + \gamma_3)(t''-t')/2} \right] .
\end{align}
Combining both results we have
\begin{align}\label{eq:two-time-mode1}
    \langle \hat{a}_1^\dagger(t'') \hat{a}_1(t') \rangle_{\rm ss} &= 
     \left( \frac{2g}{\gamma_T} \right)^2 \left\{ \frac{\gamma_T}{\gamma_1} {\rm e}^{-\gamma_1 (t''-t')/ 2} + \frac{\gamma_T}{\gamma_2 + \gamma_3 - \gamma_1}\left[ {\rm e}^{- \gamma_1 (t''-t')/2} - {\rm e}^{-(\gamma_2 + \gamma_3)(t''-t')/2} \right] \right\} + \mathcal{O}(\lambda^3) .
\end{align}
Similar expressions hold for the other two modes.

From here, the filtered output mode correlations are obtained via integration
\begin{align}
    \langle \hat{N}_i \rangle &= \gamma_{{\rm ext}, i } \int_0^\infty {\rm d}t'' \, \int_0^\infty {\rm d}t' \, f(t'') \,f(t') \langle \hat{a}^\dagger_{i} (t'') \hat{a}_{i} (t') \rangle_{\rm ss} \\
    &= \gamma_{{\rm ext}, i } \int_0^\infty {\rm d}t'' \, \int_0^{t''} {\rm d}t' \, f(t'') \,f(t') \langle \hat{a}^\dagger_{i} (t'') \hat{a}_{i}(t') \rangle_{\rm ss} \nonumber\\
    &+ \gamma_{{\rm ext}, i } \int_0^\infty {\rm d}t'' \, \int_0^{t''} {\rm d}t' \, f(t') \,f(t'') \langle \hat{a}^\dagger_{i} (t') \hat{a}_{i} (t'') \rangle_{\rm ss} ,
\end{align}
where we have split the integration up by time orderings which must be treated separately, and where we have considered a real-valued function $f(t)$.
The time antiordered correlator is defined as~\cite{Gardiner1985}
\begin{align}
    \langle \hat{a}^\dagger_{i} (t') \hat{a}_{i} (t'') \rangle_{\rm ss} = {\rm tr} \left[ \hat{a}_i {\rm e}^{\mathcal{L} (t'' - t')} \left(  \hat{\rho}_{\rm ss} \hat{a}^\dagger_i \right) \right] .
\end{align}
It is straightforward to show that the time antiordered correlator gives an identical contribution to \eqref{eq:two-time-mode1}.

Finally, we can write a general expression for the mean-number of photons in the filtered output of mode 1
\begin{align}
    \langle \hat{N}_1 \rangle &= 2 \gamma_{{\rm ext},1} \left( \frac{2g}{\gamma_T} \right)^2 \int_0^\infty {\rm d}t'' \, \int_0^{t''} {\rm d}t' \, f(t'') \,f(t') \, \langle \hat{a}^\dagger_{1} (t'') \hat{a}_{1} (t') \rangle_{\rm ss} ,
\end{align}
with the ordered two-time correlator given by \eqref{eq:two-time-mode1}.
The factor of 2 on the right hand side results from both
the time-ordered and the time antiordered contributions being identical to each other.
As the filter function $f$ is independent of $g$, the only effect of the integration is to alter the prefactor in front of $g^2$ as compared to the corresponding relation for the cavity field. 
Thus, the quadratic scaling in $g$ holds for both the cavity, and the filtered output field. 

\subsection{Steady-state output field three-time correlations}

Three-time output field correlations are much simpler to calculate as the output annihilation operators commute with each other. 
The lowest order contribution to the time-ordered correlator $\langle \hat{a}_{{\rm out},k}(t''') \hat{a}_{{\rm out},j}(t'') \hat{a}_{{\rm out},i}(t') \rangle_{\rm ss}$  ($t'''>t''>t'$, $i \neq j \neq k$) 
results from evolving the approximated steady-state with the free propagator $V_0(t)$. We use the indices $i$, $j$ and $k$ to denote an arbitrary ordering of the output operators. 
Using relations \eqref{eq:liouvillian-relation1} and \eqref{eq:liouvillian-relation2} we find that
\begin{align}
    \langle \hat{a}_{{\rm out},k}(t''') \hat{a}_{{\rm out},j}(t'') \hat{a}_{{\rm out},i}(t') \rangle_{\rm ss} &= \sqrt{\gamma_{{\rm ext},1}\, \gamma_{{\rm ext},2}\, \gamma_{{\rm ext},3}} \, \langle \hat{a}_k(t''') \hat{a}_j(t'') \hat{a}_i(t') \rangle_{\rm ss}\\
    &= \sqrt{\gamma_{{\rm ext},1}\, \gamma_{{\rm ext},2}\, \gamma_{{\rm ext},3}} \,\, {\rm tr} \left\{ \hat{a}_k \left\{ V_0(t'''-t'') \left\{ \hat{a}_j \left[ V_0 (t''-t') \left( \hat{a}_i \hat{\rho}_{\rm ss} \right) \right] \right\} \right\} \right\} \\ 
    &= 
    \sqrt{\gamma_{{\rm ext},1}\, \gamma_{{\rm ext},2}\, \gamma_{{\rm ext},3}} \, \left( \frac{-2ig}{\gamma_T} \right) {\rm e}^{-(\gamma_j + \gamma_k)(t''-t')/2} {\rm e}^{-\gamma_k (t'''-t'')/2} .
\end{align}
In this case, the only nonzero contribution comes from the first-order correction to the steady-state $-i\lambda \ket{111}\bra{000}$.

To calculate the filtered output field three-photon correlator
\begin{align}
    \langle \hat{A}_3 \hat{A}_2 \hat{A}_1 \rangle 
    &= \int_0^\infty {\rm d}t' \, {\rm d}t'' \, {\rm d}t''' \, f(t') f(t'') f(t''') 
    \langle \hat{a}_{{\rm out},3}(t''') \hat{a}_{{\rm out},2}(t'') \hat{a}_{{\rm out},1}(t') \rangle_{\rm ss} ,
\end{align}
we need to consider all of the possible time orderings inside of the integral. Without loss of generality we can fix $t'''>t''>t'$, and split the total integral into a sum of $3!$ terms corresponding to 
all possible permutations of $t'$, $t''$ and $t'''$.
Once we have assigned a time variable to each output operator $\hat{a}_{{\rm out},i}$ we can time-order them, that is, arrange them from right to left 
in order of increasing time using the fact that the output operators commute with each other. The resulting expression for the three-photon correlator is
\begin{align}
    \langle \hat{A}_3 \hat{A}_2 \hat{A}_1 \rangle &= \sum_{{\rm perm}\{i,j,k\}} \int_0^\infty {\rm d}t''' \, \int_0^{t'''} {\rm d}t'' \int_0^{t''} \, {\rm d}t' \, f(t') f(t'') f(t''') 
    \langle \hat{a}_{{\rm out},k}(t''') \hat{a}_{{\rm out},j}(t'') \hat{a}_{{\rm out},i}(t') \rangle_{\rm ss} ,
\end{align}
where we sum over all possible permutations of the mode indices $1$, $2$ and $3$.

\subsection{Steady-state output field four-time correlations}

As an example, 
let us directly consider the calculation of the number-number correlator $\langle \hat{N}_2 \hat{N}_3 \rangle$
\begin{align}
    \langle \hat{N}_2 \hat{N}_3 \rangle &= \int_0^\infty {\rm d}t_4 \, {\rm d}t_3 \, {\rm d}t_2 \, {\rm d}t_1 \, f^*(t_4) f(t_3) f^*(t_2) f(t_1)
    \langle \hat{a}^\dagger_{{\rm out},2}(t_4) \hat{a}_{{\rm out},2}(t_3) \hat{a}^\dagger_{{\rm out},3}(t_2) \hat{a}_{{\rm out},3}(t_1) \rangle_{\rm ss} ,\nonumber\\
    &= \int_0^\infty {\rm d}t_4 \, {\rm d}t_3 \, {\rm d}t_2 \, {\rm d}t_1 \, f(t_4) f(t_3) f(t_2) f(t_1)
    \langle \hat{a}^\dagger_{{\rm out},2}(t_4) \hat{a}^\dagger_{{\rm out},3}(t_2)\hat{a}_{{\rm out},2}(t_3) \hat{a}_{{\rm out},3}(t_1) \rangle_{\rm ss} ,
\end{align}
where the second line follows from the fact that field operators for different modes commute with each other, and we are restricting to real-valued filter functions $f$. From here on, we follow the prescription introduced in Ref.~\cite{Gardiner1985} to calculate the multi-time correlator; we consider a normally ordered product of time-ordered annihilation operators and time antiordered creation operators. This is always possible to do as annihilation (creation) operators for different modes commute.  
This means we can always bring any four-time correlator to the following form~\cite{Gardiner1985}
\begin{align}
    \langle \hat{B}_1(r_1) \hat{B}_2(r_2) \hat{C}_2(s_2) \hat{C}_1(s_1) \rangle 
\end{align}
with $r_2 > r_1$ and $s_2 > s_1$. Here the $\hat{B}$ operators represent creation operators, and the $\hat{C}$ operators represent annihilation operators. To solve this, we first time order the set $\{ r_1, r_2, s_1, s_2 \}$. Let us denote the time ordered elements $z_l$, with $l=1,2,3,4$ and $z_4 > z_3 > z_2 >z_1$. 
Next, we define the superoperators $\hat{F}_l$ acting at time $z_l$ as
\begin{align}
    \hat{F}_l [\hat{O}] &= \hat{C}_i \hat{O} ,\,\,\, {\rm if} \,\,\, z_l = s_i \\
    \hat{F}_l [\hat{O}] &= \hat{O} \hat{B}_j ,\,\,\, {\rm if} \,\,\, z_l = r_j , 
\end{align}
where $\hat{O}$ corresponds to an arbitrary operator. Finally, we have
\begin{align}\label{eq:general-four-time}
    \langle \hat{B}_1(r_1) \hat{B}_2(r_2) \hat{C}_2(s_2) \hat{C}_1(s_1) \rangle &= {\rm tr} \left\{ \hat{F}_4 V(t_4-t_3) \hat{F}_3 V(t_3-t_2) \hat{F}_2 V(t_2-t_1) \hat{F}_1 \hat{\rho}_{\rm ss} \right\} .
\end{align}
Here we restrict to the lowest order ($\lambda^2$) terms contributing to this correlator. These contributions arise from the terms $\beta \ket{111}\bra{111} + \epsilon \ket{011} \bra{011}$, and $i ( 2g/\gamma_T ) \left( \ket{000}\bra{111} - \ket{111} \bra{000} \right)$ in our steady-state approximation.
Starting from $\beta \ket{111}\bra{111} + \epsilon \ket{011} \bra{011}$, a nonzero contribution is obtained through the evolution with $V_0(t)$ solely, i.e., replacing the four $V$ propagators in \eqref{eq:general-four-time} by $V_0$. 
Fixing $t_4 > t_3 > t_2 > t_1$, the sum over all possible $4!$ permutations of the time indices is given by 
\begin{align}\label{eq:four-time0}
    \sum_{{\rm perm} \{i,j,k,l \}}  \langle \hat{a}^\dagger_{2}(t_i) \hat{a}_{2}(t_j) \hat{a}^\dagger_{3}(t_k) \hat{a}_{3}(t_l) \rangle_{\rm ss}    &= 4\left( \frac{2g}{\gamma_T} \right)^2 \frac{\gamma_T}{\gamma_2 + \gamma_3} \nonumber\\
    &\times \left[ {\rm e}^{\gamma_2 t_1 - 0.5 \gamma_2 t_3 - 0.5 \gamma_2 t_4 + 0.5 \gamma_3 t_1 - 0.5 \gamma_3 t_2} + {\rm e}^{\gamma_2 t_1 -0.5 \gamma_2 t_2 -0.5\gamma_2 t_4 + 0.5 \gamma_3 t_1 - 0.5 \gamma_3 t_3}
    \right.\nonumber\\
    &+ \left. {\rm e}^{0.5 \gamma_2 t_1 - 0.5 \gamma_2 t_4 + \gamma_3 t_1 - 0.5 \gamma_3 t_2 - 0.5 \gamma_3 t_3} + {\rm e}^{\gamma_2 t_1 - 0.5 \gamma_2 t_2 - 0.5 \gamma_2 t_3 + 0.5 \gamma_3 t_1 - 0.5 \gamma_3 t_4} \right.\nonumber\\
    &+ \left. {\rm e}^{0.5 \gamma_2 t_1 - 0.5 \gamma_2 t_3 + \gamma_3 t_1 - 0.5 \gamma_3 t_2 - 0.5 \gamma_3 t_4} + {\rm e}^{0.5 \gamma_2 t_1 - 0.5 \gamma_2 t_2 + \gamma_3 t_1 - 0.5 \gamma_3 t_3 - 0.5 \gamma_3 t_4} \right] .
\end{align}
On the other hand, starting from $i ( 2g/\gamma_T ) \left( \ket{000}\bra{111} - \ket{111} \bra{000} \right)$, a nonzero correction is obtained by replacing a single one of the $V$ propagators in \eqref{eq:general-four-time} by $V_1$ and the other three by $V_0$. 
Replacing $V(t_2-t_1)$ by $V_1(t_2-t_1)$ in \eqref{eq:general-four-time}, and taking into account all possible time indices permutations yields
\begin{align}\label{eq:four-time1-1}
    \sum_{{\rm perm} \{i,j,k,l \}}  \langle \hat{a}^\dagger_{2}(t_i) \hat{a}_{2}(t_j) \hat{a}^\dagger_{3}(t_k) \hat{a}_{3}(t_l) \rangle_{\rm ss}  &= -2\left( \frac{2g}{\gamma_T} \right)^2\frac{\gamma_T}{\gamma_1 - \gamma_2 - \gamma_3}   
    \left\{ {\rm e}^{0.5[-\gamma_1(t_2 - t_1)-\gamma_2 (t_4 + t_3 - t_2 - t_1)  ]} 
     \right.\nonumber\\
    &- {\rm e}^{0.5[-\gamma_3 (t_2-t_1) - \gamma_2(t_4 + t_3 - 2t_1)]} - \left. {\rm e}^{0.5[-\gamma_2 (t_4 + t_2 - 2t_1) - \gamma_3 (t_3-t_1)]} \right. \nonumber\\
    &+ \left. {\rm e}^{0.5 [-\gamma_1 (t_2 - t_1) - \gamma_2 (t_4 - t_2) -\gamma_3 (t_3 - t_1)]} - {\rm e}^{0.5[-\gamma_2 (t_4-t_1)-\gamma_3 (t_3 + t_2 - 2t_1)]}  \right. \nonumber\\
    &+ \left. {\rm e}^{0.5[-\gamma_1 (t_2 - t_1) - \gamma_2 (t_4-t_1) - \gamma_3 (t_3 - t_2)]} +  {\rm e}^{0.5[-\gamma_2 (t_3 + t_2 - 2t_1)-\gamma_3 (t_4 - t_1)]} \right. \nonumber\\
    &+ \left. {\rm e}^{0.5[-\gamma_1 (t_2 - t_1) - \gamma_2 (t_3-t_2) - \gamma_3(t_4-t_1)]} - {\rm e}^{0.5[-\gamma_2 (t_3 - t_1) - \gamma_3 (t_4 + t_2 - 2t_1)]}  \right.\nonumber\\
    &+ \left.  {\rm e}^{0.5[-\gamma_1(t_2-t_1)-\gamma_2(t_3-t_1)-\gamma_3(t_4-t_2)]} - {\rm e}^{0.5[-\gamma_2(t_2-t_1)-\gamma_3(t_4+t_3-2t_1)]} \right. \nonumber\\
    &+ \left. {\rm e}^{0.5[-\gamma_1(t_2-t_1)-\gamma_3(t_4+t_3-t_2-t_1)]} 
    \right\} .
\end{align}
Similarly, replacing $V(t_3-t_2)$ by $V_1(t_3-t_2)$ in \eqref{eq:general-four-time} yields
\begin{align}\label{eq:four-time1-2}
    \sum_{{\rm perm} \{i,j,k,l \}}  \langle \hat{a}^\dagger_{2}(t_i) \hat{a}_{2}(t_j) \hat{a}^\dagger_{3}(t_k) \hat{a}_{3}(t_l) \rangle_{\rm ss} &= -\left( \frac{2g}{\gamma_T} \right)^2\frac{\gamma_T \,{\rm e}^{-\gamma_1 (t_2 - t_1)/2}}{\gamma_1 - \gamma_2 - \gamma_3}  \left[ {\rm e}^{-\gamma_T (t_3 - t_2)/2} - 1 \right]\nonumber\\
    &\times \left[ {\rm e}^{0.5 (\gamma_2 t_2 + \gamma_3 t_1)} + {\rm e}^{0.5 (\gamma_2 t_1 + \gamma_3 t_2)} \right] \left[ {\rm e}^{-0.5(\gamma_2 t_4 + \gamma_3 t_3)} + {\rm e}^{-0.5(\gamma_2 t_3 + \gamma_3 t_4)} \right] .
\end{align}
Finally, it is apparent from the structure of the initial state that the contribution obtained from replacing $V(t_4-t_3)$ by $V_1(t_4-t_3)$ is identical to zero. 
Therefore, 
the lowest order contribution to the $\langle \hat{N}_2 \hat{N}_3 \rangle$ correlator is obtained by integrating
the sum of \eqref{eq:four-time0}, \eqref{eq:four-time1-1} and \eqref{eq:four-time1-2}, weighed by the product of the filter function $f$ at different times for $t_4 > t_3 > t_2 > t_1$. 

\section{Numerical methods - input-output with quantum pulses}\label{section:numerical_methods}

In Ref.~\cite{Kiilerich2019} the authors introduced a quantum cascaded system approach to describe the dynamics of a quantum system being driven by a continuum of modes and/or emitting radiation into a continuum of modes in a given temporal envelope or wave-packet. We briefly review the theory behind this method below. In the case of our experimental setup, the quantum system corresponds to the nonlinear microwave resonator, the input port corresponds to the flux line and the output port to the readout transmission line.

In our numerical treatment we neglect the coupling to the flux line and describe the effect of the continuous modulation of the SQUID loop on the resonator modes using Hamiltonian (2) in the main text.

First, we review the input-output with quantum pulses formalism for the case of a single resonator mode emitting radiation into a single wave-packet mode. 

\subsection{Single-mode case}

In order to describe the emission of the continuously driven microwave resonator in an arbitrary wave-packet $v(t)$, we introduce a \textit{fictitious} linear resonator with a time-dependent coupling strength $g_v(t)$ to the transmission line. 
This fictitious resonator permits us to describe the dynamics of the wave-packet mode propagating through the waveguide using a zero-dimensional quantum system. As we will detail later, choosing the correct time-dependence for the coupling strength guarantees that the propagating wave-packet is absorbed by the fictitious resonator. 

We denote by $\hat{a}$ ($\hat{a}^\dagger$) the annihilation (creation) operator of the \textit{physical} resonator. On the other hand, we denote $\hat{a}_v$ ($\hat{a}^\dagger_v$) the annihilation (creation) operator of the \textit{fictitious} resonator.
Now, both modes $\hat{a}$ and $\hat{a}_v$ are coupled to the same transmission line with strengths $\sqrt{\gamma_{\rm ext}}$ and $g_v(t)$ respectively according to the interaction Hamiltonian in the Born-Markov approximation
\begin{align}\label{app:BM_hamiltonian}
    \hat{H}_{\rm int}(t) &= i \sqrt{\frac{\gamma_{\rm ext}}{2\pi}} \int {\rm d}\omega \, \left[ \hat{a} \, \hat{b}^\dagger(\omega) \, {\rm e}^{+i \omega t} - {\rm h.c.} \right] \nonumber\\
    &+ i \frac{g_v(t)}{\sqrt{2\pi}} \int {\rm d}\omega \, \left[ \hat{a}_v \hat{b}^\dagger(\omega) \, {\rm e}^{+i \omega t} - {\rm h.c.} \right] ,
\end{align}
with $\hat{b}(\omega)$ [$\hat{b}^\dagger(\omega)$] the bosonic annihilation (creation) operators of the continuous transmission line field. 
The situation in which the output from the physical resonator is fed unidirectionally into the fictitious one is described using the formalism of cascaded systems~\cite{Combes2017}. The goal is to eliminate the degrees of freedom of the transmission line and derive an effective Lindblad-Gorini-Kossakowski-Sudarshan (LGKS) quantum master equation describing the dynamics of two coupled resonators. 
In the limit of negligible delay times, the dynamics of the two resonator modes is described by the following effective Hamiltonian 
\begin{align}\label{app:BM-Hamil}
    \hat{H}_{\rm eff}(t) &= \hat{H}_r + \frac{i \sqrt{\gamma_{\rm ext}} }{2} \left[ g_v(t) \hat{a}^\dagger_v \hat{a} - g_v^*(t) \hat{a}_v \hat{a}^\dagger \right] , 
\end{align}
and the effective decay operator
\begin{align}
    \hat{L}_{\rm eff}(t) = \sqrt{\gamma_{\rm ext}} \hat{a} + g_v(t) \hat{a}_v
\end{align}
with $\hat{H}_r$, the Hamiltonian of the physical resonator. We purposefully neglect the coherent dynamics of the fictitious resonator. Thus, the dynamics of the total system is described by the LGKS equation
\begin{align}\label{LGKS_master}
    \partial_t \hat{\rho} &= -i[\hat{H}_{\rm eff}(t), \hat{\rho}] + \mathcal{D}[\hat{L}_{\rm eff}(t)] \hat{\rho} + \sum_i \mathcal{D}[\hat{L}_i] \hat{\rho} ,
\end{align}
with $\mathcal{D}[\hat{O}] \cdot = \hat{O} \cdot \hat{O}^\dagger - \frac{1}{2} \hat{O}^\dagger \hat{O} \cdot - \frac{1}{2} \cdot \hat{O}^\dagger \hat{O}$ the dissipator superoperator. In (\ref{LGKS_master}) we have also included other possible decoherence processes (e.g. thermal gain, photon-number dephasing, etc.) described by the $\hat{L}_i$ operators. 

The last step of the derivation deals with finding the temporal profile for the coupling strength $g_v(t)$. To do this, we consider only the fictitious resonator coupled to the transmission line, i.e., the second line in \eqref{app:BM_hamiltonian}. 
The time evolution of the fictitious resonator field operator is given by
\begin{align}\label{app:av_evolution}
    \hat{a}_v(t) &= \exp \left[ -\frac{1}{2} \int_0^t {\rm d}t' \, \vert g_v(t') \vert^2 \right] \hat{a}_v(0) \nonumber\\
    & + \int_0^t {\rm d}t' \, \left\{ - g_v(t') \exp \left[ -\frac{1}{2} \int_{t'}^t {\rm d}t'' \, \vert g_v(t'') \vert^2  \right] \right\} \hat{b}_{\rm in}(t') .
\end{align}
The operator $\hat{b}_{\rm in}$ corresponds to the output field from the \textit{physical} resonator which, at the same time, corresponds to the input field to the fictitious resonator.
We want to find $g_v(t)$ such that  $\hat{a}_v(t \to \infty) = \int_0^\infty {\rm d}t' \, v^*(t') \hat{b}_{\rm in}(t')$. 

This will be the case provided that the first term on the right hand side of \eqref{app:av_evolution} vanishes in the $t \to \infty$ limit, and that
\begin{align}\label{app:v-integral}
    v^*(t) &= -g_v(t) \exp \left[ -\frac{1}{2} \int_{t}^\infty {\rm d}t' \, \vert g_v(t') \vert^2  \right], 
\end{align}
which follows from the term in curly braces on the second line of \eqref{app:av_evolution}. Our starting point is Eq.~\eqref{app:v-integral}. In order to solve for $g_v(t)$ as a function of $v(t)$ we will make the following assumptions
\begin{align}
    \vert g_v(t) \vert^2 dt &= \frac{d F(t)}{F(t)}, \label{app:cond1} \\
    F(t \to  \infty) &= 1 . \label{app:cond2}
\end{align}
From these it follows that 
\begin{align}
    \exp \left[ -\frac{1}{2} \int_t^\infty {\rm d}t' \, \vert g_v(t') \vert^2 \right] = \sqrt{F(t)} .
\end{align}
Thus, from \eqref{app:v-integral} we have $g_v(t) = -v^*(t) / \sqrt{F(t)}$. In order to satisfy the conditions \eqref{app:cond1} and \eqref{app:cond2}, we choose $F(t) = \int_0^t {\rm d}t' \, \vert v(t') \vert^2$. 
In this way, we arrive at
\begin{align}
    g_v(t) = -\frac{v^*(t)}{\sqrt{\int_0^t {\rm d}t' \, \vert v(t') \vert^2}} .
\end{align}
Finally, notice that $F(0)=0$. This guarantees that the first term on the right hand side of \eqref{app:av_evolution} vanishes as $t \to \infty$.

\subsection{Multi-mode case}

The situation we envision is the following. Three resonator modes with bosonic annihilation (creation) operators $\hat{a}_j$ ($\hat{a}^\dagger_j$) interact via the Hamiltonian $\hat{H}_{r}$. Each of these modes couples with strength $\sqrt{\gamma_{{\rm ext}, j}}$ to a transmission line. As the propagating fields at different frequencies do not interact with each other we can treat them independently. In other words, we introduce one individual environment for each resonator mode. Each of these environments will in turn, couple to a virtual cavity with time-dependent strength $g_{v,j}(t)$. These virtual cavities will store the corresponding propagating states in the wave-packet functions $v_j(t)$. This situation is described by the following interaction Hamiltonian 
\begin{align}
    \hat{H}_{\rm int}(t) &= i \sum_{j=1}^3\sqrt{\frac{\gamma_{{\rm ext}, j}}{2\pi}} \int {\rm d}\omega \, \left[ \hat{a}_j \, \hat{b}_j^\dagger(\omega) \, {\rm e}^{+i \omega t} - {\rm h.c.} \right] \nonumber\\
    &+ i \sum_{j=1}^3 \frac{g_{v,j}(t)}{\sqrt{2\pi}} \int {\rm d}\omega \, \left[ \hat{a}_{v,j} \hat{b}_j^\dagger(\omega) \, {\rm e}^{+i \omega t} - {\rm h.c.} \right] .
\end{align}

Due to the independence of the propagating outputs, it is straightforward to generalize the single-mode case studied in the previous sub-section. 
Proceeding this way, the effective Hamiltonian becomes
\begin{align}\label{app:BM-hamiltonian2}
    \hat{H}_{\rm eff}(t) &= \hat{H}_{r} + \sum_j \frac{i \sqrt{ \gamma_{{\rm ext}, j} } }{2} \left[ g_{v,j}(t) \hat{a}^\dagger_{v,j} \hat{a}_j - g_{v,j}^*(t) \hat{a}_{v,j} \hat{a}_j^\dagger \right] .
\end{align}
On the other hand, we obtain three effective decay operators for the different output modes
\begin{align}
    \hat{L}_{{\rm eff},j}(t) = \sqrt{ \gamma_{{\rm ext}, j} } \hat{a}_j + g_{v,j}(t) \hat{a}_{v,j} .
\end{align}
Finally, the full dynamics of the three resonator modes and three output modes system is described by the LGKS master equation
\begin{align}\label{eq:full-input-output}
    \partial_t \hat{\rho} &= -i[\hat{H}_{\rm eff}(t), \hat{\rho}] + \sum_{j=1}^3 \mathcal{D}[\hat{L}_{{\rm eff},j}(t)] \hat{\rho} + \sum_i \mathcal{D}[\hat{L}_i] \hat{\rho}.
\end{align}
Once again, we have included other possible decoherence processes described by the $\hat{L}_i$ operators.
We solve this equation numerically using QuTiP~\cite{Lambert2024}.

\begin{figure}[t]
    \centering
    \includegraphics[width=1.\linewidth]{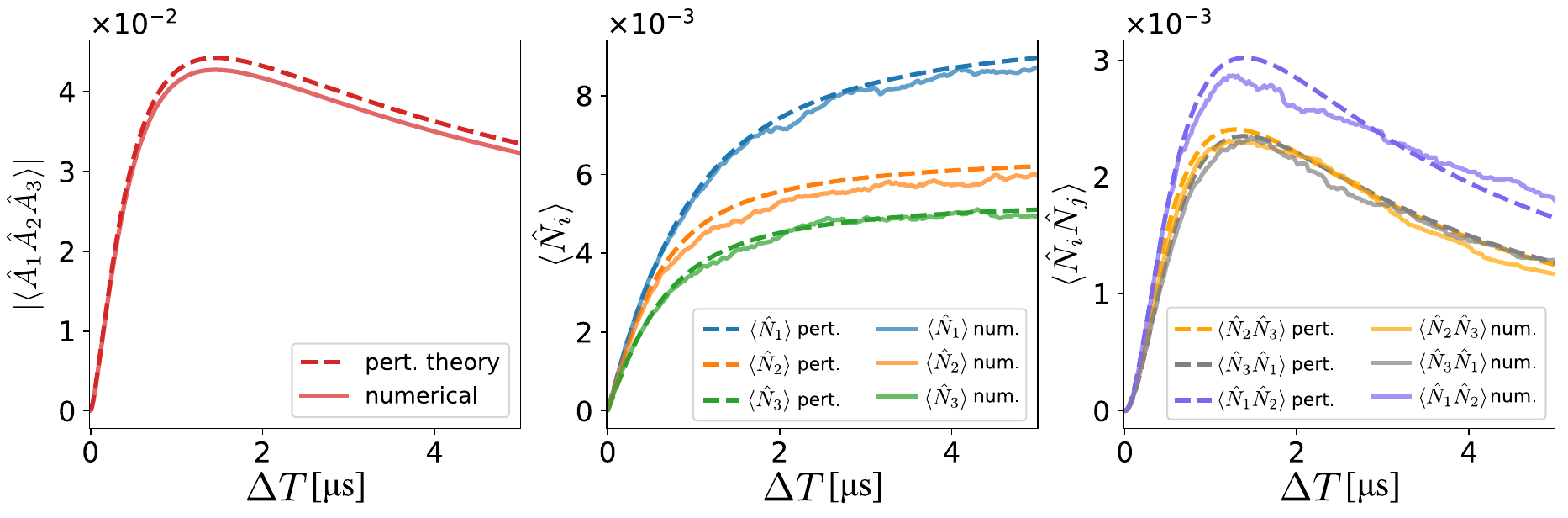}
    \caption{Filtered output field correlators as a function of the width $\Delta T$ of a boxcar filter. Analytical results following from our perturbative analysis are shown in dashed lines while numerical results following from the input-output with quantum pulses method are shown in solid lines. The choice of parameters $g/2\pi = 0.03$ MHz and $\gamma_T /2\pi = 3.032$ MHz validate the perturbative treatment. The rest of the parameters are given in the main text. }
    \label{fig:numerical-analytical}
\end{figure}

In Fig.~\ref{fig:numerical-analytical}
we calculate numerically (solid lines) the correlators $\langle \hat{A}_1 \hat{A}_2 \hat{A}_3 \rangle$, $\langle \hat{N}_i \rangle$, and $\langle \hat{N}_i \hat{N}_j \rangle$ which contribute to the entanglement witness $W$ for the case of a boxcar filter function of width $T$. 
These are obtained by numerically solving \eqref{eq:full-input-output} for 
\begin{align}
 \hat{H}_r &= g\left(\hat{a}_1\hat{a}_2\hat{a}_3+\ad_1\ad_2\ad_3\right) 
 - \sum_{n, m}\chi_{nm}\ad_n\ad_m\hat{a}_n\hat{a}_m ,
\end{align}
and considering internal and external (single-photon) losses with decay rates 
$\gamma_{{\rm int},i}$ and $\gamma_{{\rm ext},i}$ (respectively) for each mode $\hat{a}_i$ (internal losses correspond to the $\hat{L}_i$ channels in \eqref{eq:full-input-output}, i.e., $\hat{L}_i = \sqrt{\gamma_{{\text{int}}, i}} \, \hat{a}_i$).
The parameters considered in our simulations are: $g / 2\pi = 0.03$ MHz, the internal decay rates are 
$\gamma_{{\rm int},1}/ 2\pi = 0.035$ MHz, $\gamma_{{\rm int},2}/ 2\pi = 0.097$ MHz, $\gamma_{{\rm int},3}/ 2\pi = 0.308$ MHz, and the external decay rates are
$\gamma_{{\rm ext},1}/ 2\pi = 0.604$ MHz, $\gamma_{{\rm ext},2}/ 2\pi = 1.242$ MHz, $\gamma_{{\rm ext},3}/ 2\pi = 0.746$ MHz.
The Kerr and cross-Kerr strengths $\chi_{nm}$ are taken from Table~\ref{table:Kerr_Matrix} 
by considering mode 1 as the mode with frequency 5.560 GHz, mode 2 as the mode with frequency 6.831 GHz and mode 3 as the mode with frequency 7.902 GHz.

Due to the large Hilbert space corresponding to 6 bosonic modes, i.e., three resonator modes and three fictitious resonator modes, we solve the nonunitary dynamics of the system using quantum trajectories. This is the cause of the wiggly solid lines observed in the figure. For this simulation, we truncated the Hilbert space of each bosonic mode to 6 excitations, that is, a total Hilbert space dimension of $6^6$. 
In the weak-driving regime considered in this work, a very large number of trajectories is necessary in order to average out the effects of quantum fluctuations. In the example shown we have considered a total of $10^5$ trajectories.

To verify the validity of our numerical results, we compare them against analytical solutions derived using perturbation theory (dashed lines)
following Sect.~\ref{section:output}.  
We observe a very good agreement with the analytical solution. A small disagreement is expected as the analytical solution neglects Kerr nonlinearities.

\bibliography{main_SM}